\documentclass[
 reprint,
superscriptaddress,
amsmath,amssymb,
aps,
prx,
]{revtex4-2}

\usepackage{graphicx}
\usepackage{dcolumn}
\usepackage{bm}
\usepackage{hyperref}
\hypersetup{
     colorlinks   = true,
     allcolors    = blue
}
\usepackage{cleveref}
\usepackage{dsfont}
\usepackage{xcolor}
\usepackage{amsmath}
\usepackage{array}
\usepackage{braket}
\crefname{equation}{Eq.}{Eqs.}
\Crefname{equation}{Equation}{Equations}
\crefname{figure}{Fig.}{Figs.}
\Crefname{figure}{Figure}{Figures}
\crefname{section}{Sec.}{Secs.}
\Crefname{section}{Section}{Sections}
\crefname{table}{Table}{Tables}
\crefname{appsec}{}{Appendices}

\newcommand{\psiopt}{\ket{\psi^{\text{opt}}}_{x, x+1}}

\definecolor{ultramarine}{RGB}{0,32,96}

\usepackage{titlesec}
\usepackage{float}

\titleformat{\paragraph}[runin]
        {\bfseries}
        {}
        {0.0em}
        {}
        [ -- ~]
        
\titlespacing*{\paragraph}{0pt}{4pt}{0pt}

\begin{document}
\title{Multi-Target Density Matrix Renormalization Group X algorithm\\ and its application to circuit quantum electrodynamics}

\author{Sof\'ia Gonz\'alez-Garc\'ia}
\email{sofiagonzalezgarcia@ucsb.edu}
\affiliation{Google Quantum AI, Santa Barbara, CA 93111, USA}
\affiliation{Department of Physics, University of California at Santa Barbara, Santa Barbara, CA 93116, USA}

\author{Aaron Szasz}
\affiliation{Google Quantum AI, Santa Barbara, CA 93111, USA}

\author{Alice Pagano}
\affiliation{Google Quantum AI, Santa Barbara, CA 93111, USA}

\author{Dvir Kafri}
\affiliation{Google Quantum AI, Santa Barbara, CA 93111, USA}

\author{Guifr\'e Vidal}
\affiliation{Google Quantum AI, Santa Barbara, CA 93111, USA}

\author{Agustin Di Paolo}
\affiliation{Google Quantum AI, Santa Barbara, CA 93111, USA}

\begin{abstract}
Obtaining accurate representations of the eigenstates of an array of coupled superconducting qubits is a crucial step in the design of circuit quantum electrodynamics (QED)-based quantum processors. However, exact diagonalization of the device Hamiltonian is challenging for system sizes beyond tens of qubits. Here, we employ a variant of the density matrix renormalization group (DMRG) algorithm, DMRG-X, to efficiently obtain localized eigenstates of a 2D transmon array without the need to first compute lower-energy states. We also introduce MTDMRG-X, a new algorithm that combines DMRG-X with multi-target DMRG to efficiently compute excited states even in regimes with strong eigenstate hybridization. We showcase the use of these methods for the analysis of long-range couplings in a multi-transmon Hamiltonian including qubits and couplers, and we discuss eigenstate localization. These developments facilitate the design and parameter optimization of large-scale superconducting quantum processors.
\end{abstract}

\maketitle
\section{\MakeUppercase{Introduction}}

Two-dimensional arrays of coupled transmon qubits~\cite{koch_2007}, described by the theoretical framework known as circuit quantum electrodynamics (QED)~\cite{blais2004cavity,blais2021circuit}, are one of the leading platform architectures for quantum computation. Characterizing the energy spectrum of these systems is essential for their calibration, quantum control, and the design of next-generation versions. For implementing high-fidelity single- and two-qubit gates, the quantum processor should operate deep within a many-body localized (MBL) regime and far from an uncontrollable chaotic phase~\cite{berke2022transmon}. In this context, knowledge about the low-energy excitations of the 2D transmon Hamiltonian is useful to identify operating regimes---defined by the qubit energies and coupling strengths---such that unwanted couplings are minimized without sacrificing two-qubit gate speed. Additionally, with the advent of analog quantum computation, more accurate knowledge of the multi-transmon Hamiltonian is needed for Hamiltonian emulation and cross-entropy benchmarking~\cite{Lamata_2018, Bluvstein_2022, Shaw_2024, andersen2024}.

Characterizing the energy spectrum of a multi-transmon array given its Hamiltonian parameters requires working with an exponentially large state space. For this reason, doing so via exact diagonalization becomes computationally intractable for system sizes beyond tens of (multi-level) transmons. Perturbative methods, such as the Schrieffer-Wolf transformation~\cite{Schrieffer_66, Bravyi_2011}, are an excellent option for analyzing weakly interacting quantum systems, and are routinely used in circuit QED. However, perturbation theory can suffer from convergence issues when approaching the nonperturbative regime required for fast two-qubit gates, see e.g., Refs.~\cite{abanin2017rigorous,mori2018thermalization}. Thus, developing an alternative option to enable fast numerical analysis of realistic circuit models and also to serve as a reference for perturbative expressions is of great value.

\begin{table}[t!]
\centering
\begin{tabular}{c|c|c|}
 & Single target & Multi-target \\
\hline
Lowest energy & DMRG~\cite{white_1992} & MTDMRG~\cite{white_1992, Dolgov_2014, baker2021} \\
\hline
Spatial profile & DMRG-X~\cite{Khemani_2016} & MTDMRG-X (this work) \\
\hline
\end{tabular}
\caption{Variants of the DMRG algorithm given the tensor-update objective: lowest energy, or spatial profile; and number of targeted states: single target, or multiple targets. The method we introduce, MTDMRG-X, involves the targeting of multiple states based on their spatial profile.}
\label{table:methods_comparison}
\end{table}   

Tensor network methods, which have been used extensively in the field of condensed matter, provide a numerical approach for computing the excitations of many-body systems efficiently via a controlled approximation. The best known tensor network method is the density matrix renormalization group (DMRG) algorithm~\cite{white_1992}, which efficiently finds moderately-entangled ground states of interacting quantum systems.  The computed ground states are given as matrix product states (MPS)~\cite{fannes_1992, Ostlund_1995, Verstraete_2006, Verstraete_2008,  Schollwock_2011}; the matrix product state ansatz can in principle represent any quantum state on a lattice but it is only efficient for wavefunctions with limited entanglement.

DMRG and variants have been used for studying 1D superconducting circuits with multiple degrees of freedom~\cite{Chung_1997, Lee_2003, Weiss_2019, Di_Paolo_2021}. While these previous works demonstrated how useful DMRG can be for modeling circuit-QED setups, their approaches suffer from a common shortcoming: finding a target excited state of the superconducting circuit of interest requires first computing all eigenstates with lower energy than the target. While this is not an issue for systems where only a few excitations lie between the ground state and the excited state of interest (as is the case in these previous works), targeting specific excitations of a multi-transmon device in this way is impractical due to the rapid growth of the number of low-energy states with system size.

This inefficiency is partially resolved by an existing DMRG variant, DMRG-X~\cite{Khemani_2016}. DMRG-X was developed to compute excited MBL eigenstates, taking advantage of the localization of the states to target spatial structure rather than low energies.  As we demonstrate numerically in this work, DMRG-X can also be used to efficiently find target excited eigenstates of a transmon chip, without needing to find lower-energy states first.  However, DMRG-X is still inefficient when targeting not just a single localized excited state but rather a set of strongly-hybridized states.

Here, we present an adaptation of the DMRG algorithm for simultaneously targeting multiple excited states via an overlap-based tensor-update objective function, which we refer to as MTDMRG-X, see~\cref{table:methods_comparison}. As in DMRG-X~\cite{Khemani_2016}, our method allows for the targeting of excited states without needing to compute lower-energy eigenstates. However, we go beyond DMRG-X by allowing for the targeting of a subspace of excited states by combining DMRG-X with multi-target DMRG (MTDMRG)~\cite{white_1993, Chandross_1999, Degli_2004, Chepiga_2017, baker2021, Li_2023}. The resulting method is robust at resolving excited-state sets with strong hybridization and therefore ideal for analyzing multi-transmon eigenstates with a large number of excitations, see e.g. Fig. 2 in~Ref.~\cite{berke2022transmon}. By modifying the algorithm for finding eigenstates of the effective Hamiltonian, we are able to target excited states deep in the spectrum without increased computational overhead. Similarly to DMRG-X, MTDMRG-X can be run independently (and in parallel) for each target-state set of interest.

In this manuscript, we describe in detail the new tensor-network method MTDMRG-X.  We also showcase its use in practical scenarios which arise in the design of a 2D transmon quantum processor, quantifying localization of chip eigenstates corresponding to computational basis states and computing coupling-strength metrics.

\section{\MakeUppercase{How to read this paper}}

The remainder of this manuscript is organized as follows. In~\cref{sec:dmrg_background}, we review DMRG algorithms that are known in the literature and introduce relevant notation, which we use later on for describing our new algorithm.  Notably, we review not just the standard DMRG method, but also the less well-known MTDMRG and DMRG-X. In~\cref{sec:mtdmrgx}, we present MTDMRG-X, which is also our main result. In~\cref{sec:circuit QED} we introduce the circuit-QED setup of interest, and we showcase the use of both DMRG-X and MTDMRG-X algorithms for computing properties of this system: eigenstate localization and the dependence of qubit-qubit coupling on the state of other nearby qubits. We conclude in~\cref{sec:conclusions}.

Circuit-QED practitioners interested only in the capabilities of our new method as a useful ``black-box'' tool can proceed directly to~\cref{sec:circuit QED}, for sample numerical results on practical use-cases.  Those interested in also understanding some details of the numerical approach could benefit from reading all sections of the manuscript, in the order in which they are presented.  Finally, tensor network experts primarily interested in the new MTDMRG-X algorithm can go directly to~\cref{sec:mtdmrgx}, but are encouraged to at least skim~\cref{sec:dmrg_background}, where we introduce our notation.  In particular, in~\cref{sec:dmrg} we introduce a 3-step description of each two-site update during a DMRG sweep, to which we refer heavily in describing not only MTDMRG-X but also MTDMRG and DMRG-X.

\section{\MakeUppercase{Established DMRG algorithms}}
\label{sec:dmrg_background}

We first review DMRG algorithms from the literature in a way that facilitates introducing our new method, MTDMRG-X. Our review includes the standard DMRG~\cite{white_1992, white_1993}, MTDMRG~\cite{white_1993, Huang_2018, Dolgov_2014, baker2021}, and DMRG-X~\cite{Khemani_2016} algorithms. 

\subsection{Matrix Product States\label{sec:mps}}

The wavefunction of a quantum system on~$N$ sites with local dimensions~$\{d_x\}$ is of the form
\begin{equation}
    \ket{\Psi} = \sum_{\{\alpha_x\} = 1}^{d_x}\Psi_{\alpha_1,\dots,\alpha_N}\ket{\alpha_1\,\cdots\,\alpha_N},
\label{eq:many-bod_wfn_tn}
\end{equation}
where~$\Psi_{\alpha_1,\dots,\alpha_N}$ is the amplitude associated with the state configuration~$\ket{\alpha_1\,\cdots\,\alpha_N}$. This representation of the wavefunction requires storing an exponential number~$\textstyle\prod_{x=1}^Nd_x$ of coefficients. 

\begin{figure}[t!]
\centering
\includegraphics[width=0.9\columnwidth]{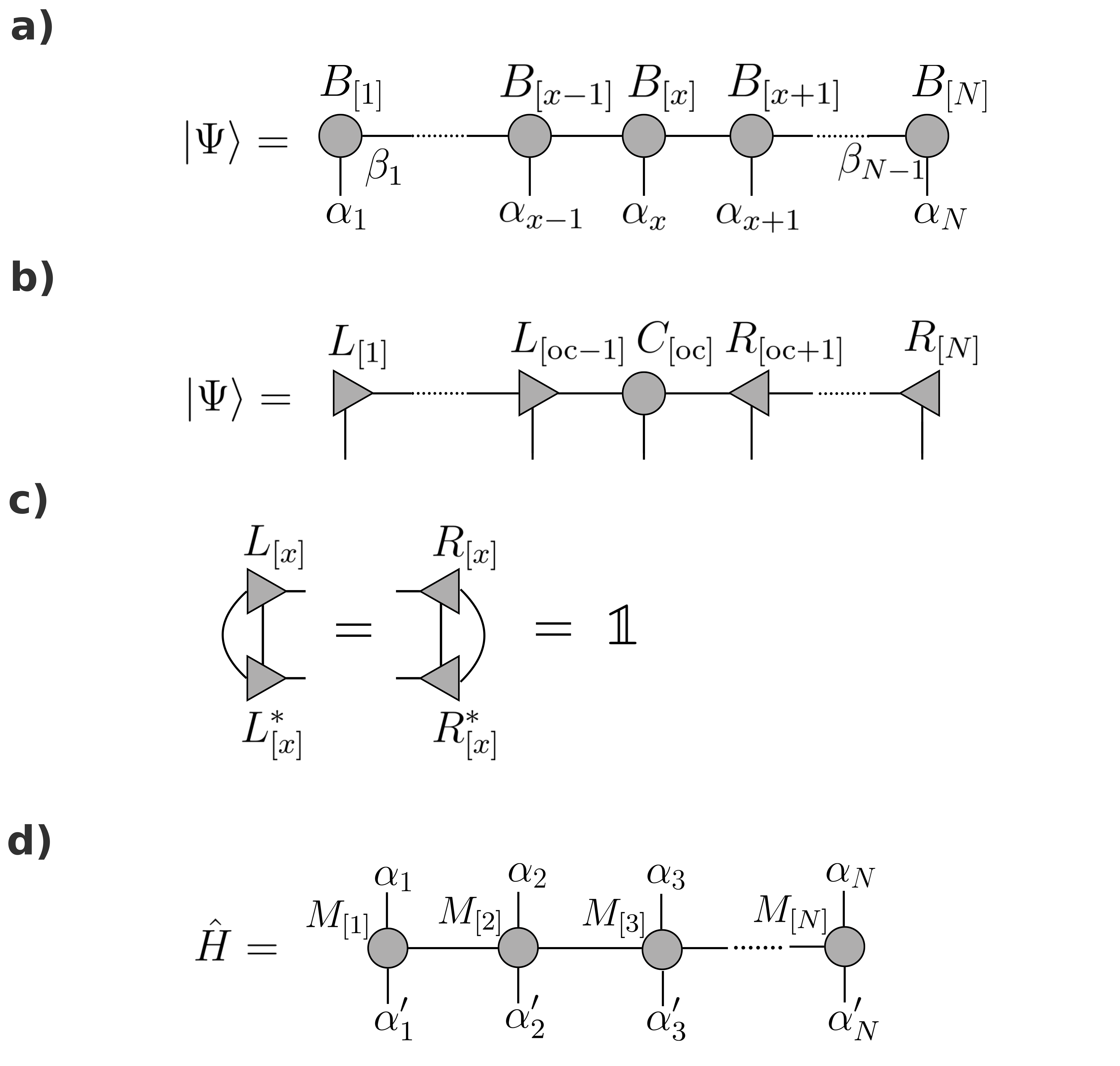}
\caption{\textbf{MPS and MPO conventions.} \textbf{\textsf{a)}} Diagrammatic depiction of the MPS ansatz. \textbf{\textsf{b)}} MPS ansatz in the center of orthogonality canonical form. \textbf{\textsf{c)}} Left and right isometric conditions. \textbf{\textsf{d)}} Matrix product operator.}
\label{fig:mps_objects} 
\end{figure}

A more efficient representation of many-body quantum states with low-to-moderate entanglement is provided by matrix product states (MPS)~\cite{white_1992, fannes_1992}.  In an MPS, the amplitudes~$\Psi_{\alpha_1,\dots,\alpha_N}$ are given by products of matrices.  
Specifically, for each site~$x$ and for each local basis state~$|\alpha\rangle \in \{|1\rangle,\cdots,|d_x\rangle\}$, we assign a matrix~$B_{[x]}^{\alpha}$.  The matrices for a given site~$x$ all have the same dimensions, $\text{dim}(B_{[x]}^{\alpha}) = \chi_{x-1} \times \chi_x$, with~$\chi_0 = \chi_N = 1$.  In this notation, the amplitudes are 
\begin{equation}
    \Psi_{\alpha_1,\dots,\alpha_N} = B_{[1]}^{\alpha_1}B_{[2]}^{\alpha_2} \cdots B_{[N-1]}^{\alpha_{N-1}} B_{[N]}^{\alpha_N}. \label{eq:mps_equation}
\end{equation}
Using standard tensor network diagrammatic notation, we show an MPS in ~\cref{fig:mps_objects}a.  Each circle represents the collection of matrices, $\{B_{[x]}^{\alpha_x}\, | \,\alpha_x \in 1,\cdots,d_x\}$, on the corresponding site.  We view this collection as a rank-3 tensor with indices~$\alpha_x,\beta_{x-1}$, and~$\beta_x$.  The ``physical index''~$\alpha_x$ corresponds to different choices of the local basis state and is indicated in the figure by vertical lines from the tensors.  The ``bond indices''~$\beta_{x-1}, \beta_x$ are the ones summed over in matrix multiplication,
\begin{align}
     &\Psi_{\alpha_1,\dots,\alpha_N} \label{eq:mps_equation_2}\\
     &= \sum_{\beta_1,\cdots, \beta_{N-1}} B_{[1]}^{\alpha_1,\beta_1} B_{[2]}^{\alpha_2,\beta_{1}\beta_{2}} \cdots B_{[N-1]}^{\alpha_{N-1},\beta_{N-2}\beta_{N-1}} B_{[N]}^{\alpha_N,\beta_{N-1}}.\nonumber
\end{align}
This tensor contraction is shown in the figure via the horizontal lines connecting the sites.
We will use the MPS as a variational ansatz for a quantum state.  
The largest dimension~$\max(\chi_x)~\forall x \in \{1,\cdots,N\}$ is referred to as the MPS bond dimension~$\chi$.  The representation power of this ansatz depends on~$\chi$, and for~$N$ sites with local dimension~$d$, the number of variational parameters scales as~$\mathcal{O}(Nd\chi^2)$, which for fixed~$\chi$ is no longer exponential in the system size.

There exists a gauge degree of freedom in the bond indices of the tensors, allowing the transformation of the MPS into specific canonical forms. It is convenient to work in a canonical form called the center of orthogonality form. Given a site~$x$ that we choose as the orthogonality center ($x=\text{oc}$) the MPS is partitioned into left and right isometric pieces whose tensors form left and right isometries, respectively, see~\cref{fig:mps_objects}c for isometry conditions. To indicate the use of this canonical form, we rename tensors satisfying the respective isometry conditions from~$B$ to either~$L$ or~$R$ and depict them as left and right pointing triangles, respectively. The tensor (or bond) separating the two isometric parts is the orthogonality center~$C_{[x=\text{oc}]}$. See~\cref{fig:mps_objects}b for a depiction of the MPS ansatz in the center of orthogonality canonical form. Mathematically:
\begin{equation}
    \ket{\Psi} = \sum_{\alpha_1,\cdots, \alpha_{N}} L_{[1]}^{\alpha_1} \cdots C_{[\text{oc}]}^{\alpha_{\text{oc}}} \cdots R_{[N]}^{\alpha_{N}} \ket{\alpha_1 \cdots \alpha_{N}},
\label{eq:mps_center_orthogonality}
\end{equation}
where as in ~\cref{eq:mps_equation}, the coefficient of each basis state is represented as a matrix product.

When working with states given in MPS form, it is helpful to represent operators, such as the Hamiltonian, as matrix product operators (MPOs).  An MPO is shown diagrammatically in~\cref{fig:mps_objects}d.  Each tensor, $M_{[x]}$, is now rank 4, with coefficients~$M_{[x]}^{\alpha_{x}\alpha_{x}'\beta_{x-1}\beta_x}$. The indices on the vertical legs, $\alpha_x$ and~$\alpha_x'$, indicate the local basis states acted on by the tensor, while the horizontal legs again indicate matrix multiplication.  To be precise, for an operator~$\hat{O}$ represented as an MPO, the matrix element~$\bra{{\alpha'}_1\cdots{\alpha'}_N}\hat{O}\ket{\alpha_1\cdots\alpha_N}$ is given by
\begin{equation}
    = \sum_{\beta_1,\cdots, \beta_{N-1}} M_{[1]}^{\alpha_1\alpha_1',\beta_1} M_{[2]}^{\alpha_2\alpha_2',\beta_{1}\beta_{2}} \cdots 
    M_{[N]}^{\alpha_N\alpha_N',\beta_{N-1}}.\label{eq:MPO_element}
\end{equation}

\subsection{Density Matrix Renormalization Group\label{sec:dmrg}}
\begin{figure}[t!]
\centering
\includegraphics[width=0.9\columnwidth]{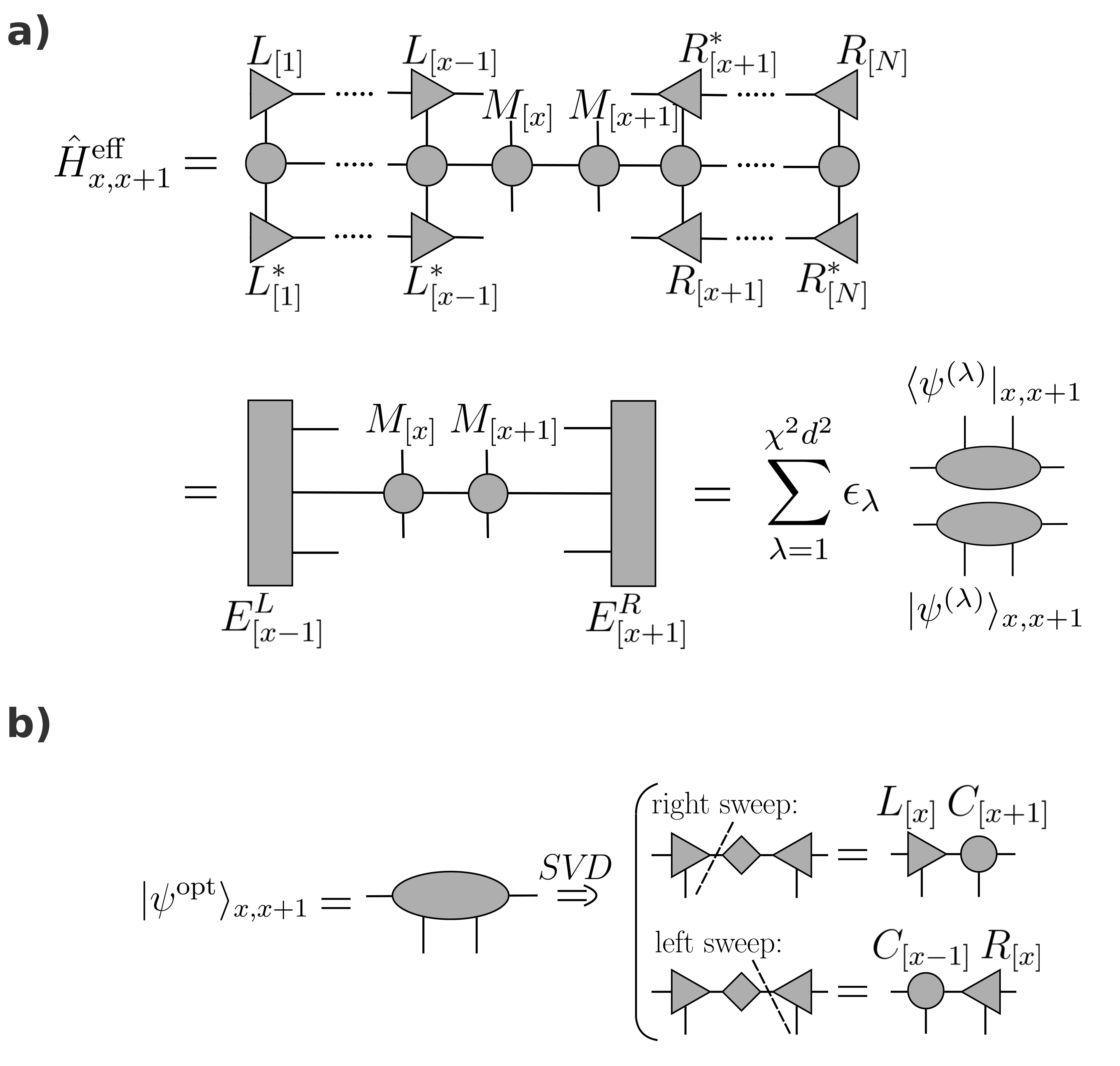}
\caption{\textbf{Two-site DMRG update.} \textbf{\textsf{a)}} Construction of the effective Hamiltonian~$\hat{H}^{\text{eff}}_{x,x+1}$ for the DMRG update of sites~$x,x+1$.  The left and right isometric parts of the MPS and the MPO are contracted into left and right environments~$E^L$ and~$E^R$. \textbf{\textsf{b)}} Decomposition of the optimal effective two-site state into MPS form for a left and right sweep.  The dashed diagonal line indicates the location of bond dimension truncation.}
\label{fig:dmrg_objects} 
\end{figure}

The MPS ansatz can be efficiently optimized to approximately represent low energy eigenstates of a gapped local Hamiltonian~$\hat{H}$ through the density matrix renormalization group (DMRG) algorithm~\cite{white_1992, Verstraete_2006, Schollwock_2011}.  
The DMRG algorithm begins by representing the Hamiltonian as an MPO, then choosing an initial state, either random or an informed guess, represented as an MPS in the center of orthogonality canonical form with the orthogonality center at the first site.
The state is then iteratively optimized by performing consecutive two-site updates organized into right and left sweeps. Below we will outline the steps of a two-site DMRG update of sites~$x,x+1$ in a right sweep.  These steps assume the MPS is in the center of orthogonality canonical form with orthogonality center in site~$x$. 
The tensors on sites~$x$ and~$x+1$ are updated by following these steps:

\begin{enumerate}
    \item[i)] \emph{Obtain an effective two-site Hamiltonian}, $\hat{H}^{\text{eff}}_{x,x+1}$, 
    by using the current state~$\ket{\Psi}$ to project the MPO for the Hamiltonian onto the bond indices~$\beta_{x-1}, \beta_{x+1}$.  This is done by computing the object~$\braket{\Psi |\hat{H}|\Psi}$ with the tensors on sites~$x,x+1$ removed from~$\bra{\Psi}$ and~$\ket{\Psi}$. See~\cref{fig:dmrg_objects}a.  The resulting~$\hat{H}^{\text{eff}}$ is viewed as acting on the space of 4-index vectors with dimensions~$(\chi_{x-1},d,d,\chi_{x+1})$. The effective Hamiltonian has eigenstate decomposition given by:
    \begin{equation}
        \hat{H}^{\text{eff}}_{x,x+1} = \sum_{\lambda=1}^{\chi^2d^2} \epsilon_{\lambda} \ket{\psi ^{(\lambda)}}_{x,x+1}\bra{\psi ^{(\lambda)}}_{x,x+1}
    \end{equation}

    \item[ii)] \label{point:choice_heff} \emph{Obtain the ground-state of}~$\hat{H}^{\text{eff}}$. To minimize the global energy expectation value, aiming at finding the global ground state, the optimal choice~$\psiopt$  for updating the local state on sites~$x$ and~$x+1$ is to find the vector that minimizes the energy of the effective Hamiltonian, namely~$\ket{\psi ^{(1)}_{x,x+1}}$.

    \item[iii)] \emph{Recover the matrix product state form} by performing a singular value decomposition (SVD) of the two-site effective tensor from the previous step (see~\cref{fig:dmrg_objects}b). At this point, the tensors from the SVD are truncated to the bond dimension~$\chi$.
    
\end{enumerate}

The optimization sites are shifted by one site~$(x, x+1)\to(x+1, x+2)$ and the procedure is repeated to continue with the right DMRG sweep. This implementation is easily generalized for a left sweep.

The DMRG terminates when the total energy measured from the effective two-site states has converged below some threshold; see~\cref{app:evaluating_convergence}. In our numerical experiments (\cref{sec:circuit QED}) the energy convergence threshold is set to be~$10^{-10}$ GHz. The accuracy of the final state depends on the value of the bond dimension~$\chi$. To assess the accuracy of an eigenstate obtained through DMRG we use the variance of the Hamiltonian expectation value:
\begin{equation}
    \text{var}(\hat{H}, \ket{\Psi}) = \braket{\Psi|\hat{H}^2|\Psi} - \braket{\Psi|\hat{H}|\Psi}^2.
\label{eq:psi_hamiltonian_variance}
\end{equation} 
Using this metric, the true targeted eigenstate~$\ket{\Tilde{\Psi}}$ would have~$\text{var}(\hat{H}, \ket{\Tilde{\Psi}})=0$. The energy error~$\Delta E = \left|\braket{\Psi| \hat{H} | \Psi} - \braket{\Tilde{\Psi}| \hat{H} | \Tilde{\Psi}}\right|$ and Hamiltonian variance~$\text{var}(\hat{H}, \ket{\Psi})$ both depend linearly on the state infidelity~$\varepsilon = 1- \left|\braket{\Psi | \Tilde{\Psi}}\right|^2$. See~\cref{app:mps_accuracy}. Therefore the Hamiltonian variance is informative of the energy error for a given MPS bond dimension.

\subsubsection{Solving for excited states with DMRG\label{sec:dmrg_es}}

In addition to finding the ground state, DMRG and related algorithms can also be used to find excited states.  Here we review three approaches: orthogonalizing against previously found lower-energy states~\cite{Stoudenmire_2012}; multi-target DMRG~\cite{white_1993, Huang_2018, Dolgov_2014, baker2021}, where we use a modified MPS to target multiple low-energy states simultaneously; and DMRG-X~\cite{Khemani_2016}, where we use known spatial structure of wavefunctions to target specific excited eigenstates. 
\\

\textit{DMRG with orthogonalization---}The most common method for obtaining excited states is to project out the lower energy eigenstates in the Hamiltonian. The modified Hamiltonian~$\hat{H}^{'(m)}$ for finding excited eigenstate~$m$ after having obtained eigenstates~$\{\ket{\Psi_1}, \dots, \ket{\Psi_{m-1}}\}$ will be: 
\begin{equation}
     \hat{H}^{'(m)} = \left( \mathds{1} - \sum_{j=1}^{m} \ket{\Psi_j}\bra{\Psi_j}\right)  \hat{H} \left( \mathds{1} - \sum_{j=1}^{m} \ket{\Psi_j}\bra{\Psi_j}\right).
     \label{eq:H_ortho}
\end{equation}
Under~$\hat{H}^{'(m)}$, all previously found states have zero energy, so an excited state of~$\hat{H}$ becomes the new ground state and can be found using DMRG. 
In practice, we do not modify the full Hamiltonian via its MPO representation.  Instead, we only need to modify step ii) of the two-site update in the DMRG algorithm.  Specifically, we find~$\psiopt$ that minimizes~$\tilde{H}^\text{eff}_{x,x+1}$, which is~$H^\text{eff}_{x,x+1}$ projected into the space orthogonal to the previously found eigenstates.

This method requires obtaining excited states one by one. If we are only interested in the~$m^{\text{th}}$ state, we must still first find all the lower energy states~$\{\ket{\psi_j}\}$ for~$j \in \{1,\dots,m-1\}$. This quickly becomes computationally expensive; to find~$m$ states, the cost scales as~$m^2$.   Furthermore, convergence across excited states is not uniform because an error in one state will affect the effective Hamiltonian used to find subsequent states, so that errors accumulate as more excited states are found.\\

\textit{Multi-target DMRG---}To avoid the problem of compounding error in subsequent eigenstates, we can instead use a variation of the DMRG algorithm to target the lowest~$m$ eigenstates all simultaneously, with nearly uniform convergence across the different eigenstates. The idea of simultaneously obtaining the groundstate and multiple states was already mentioned by White in~\cite{white_1993}. There exist several subsequent algorithms in the literature that demonstrate successful multiple excited state convergence beyond the standard DMRG approach~\cite{Chandross_1999, Degli_2004,Dolgov_2014, Chepiga_2017,Huang_2018, baker2021, Li_2023}. Here we will briefly review the multi-target MPS (MTMPS) ansatz and multi-target DMRG (MTDMRG) approach as outlined in~\cite{baker2021}.

\begin{figure}[t!]
\centering
\includegraphics[width=0.95\columnwidth]{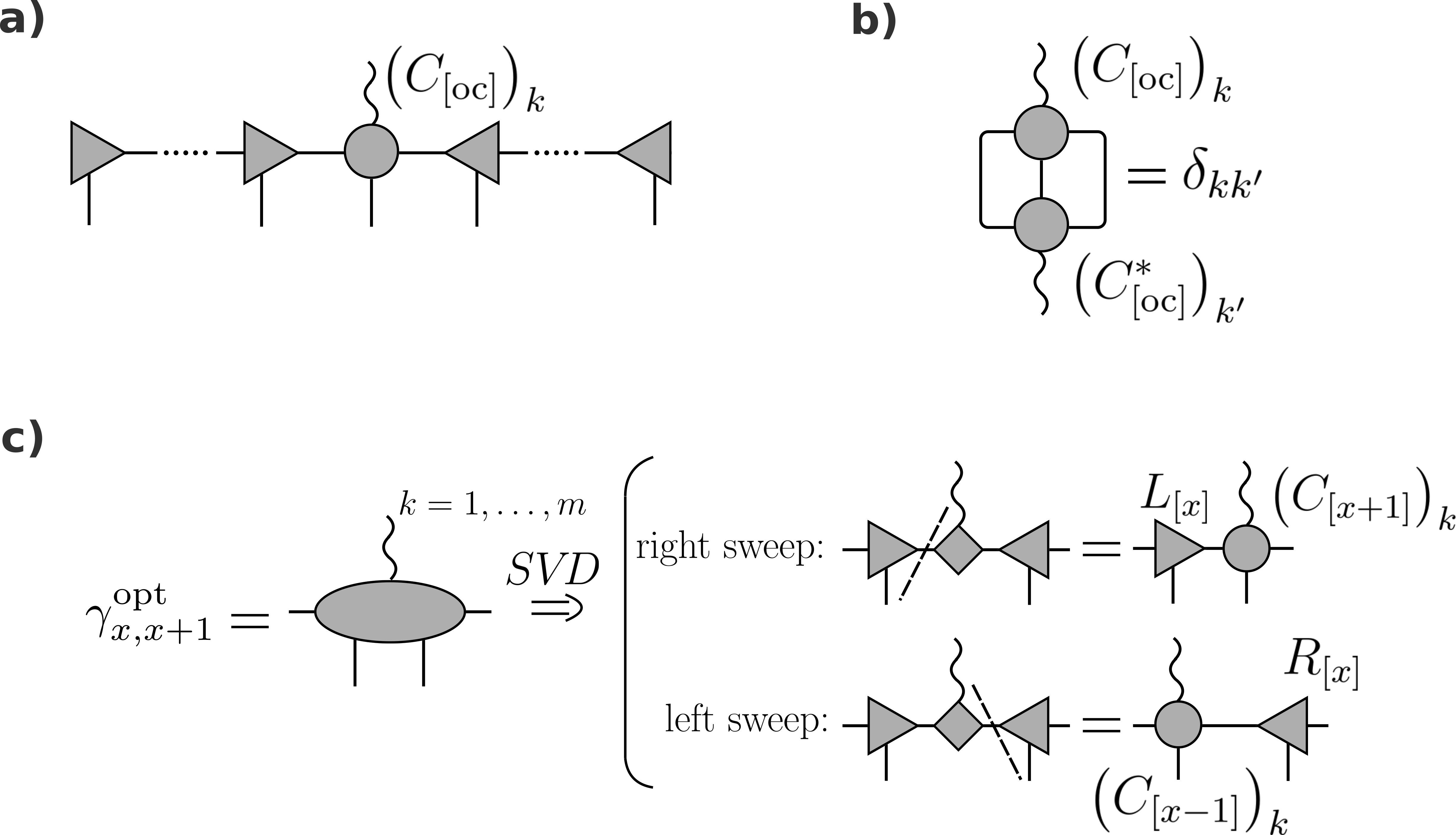}
\caption{\textbf{Multi-target DMRG adaptations.} \textbf{\textsf{a)}} Multi-target MPS ansatz, with the multi-target index attached to the orthogonality tensor~$C_{\text{oc}}$. \textbf{\textsf{b)}} MTMPS orthonormality condition. \textbf{\textsf{c)}} Decomposition of the optimal multi-target effective state into MTMPS form for a left and right sweep update. The dashed diagonal line indicates the location of bond dimension truncation.}
\label{fig:mtdmrg_diags} 
\end{figure}

Using the center of orthogonality canonical form, the MPS ansatz can be generalized to represent~$m$ quantum states by attaching a multi-target index~$k$ to the tensor in the orthogonality center~$C_{[\text{oc}]}$. Diagrammatically, we represent this as a squiggly leg, see~\cref{fig:mtdmrg_diags}a. This ansatz~$\Gamma$ is referred to as the multi-target MPS (MTMPS), and mathematically it is:

\begin{equation}
    \Gamma = \sum_{k=1}^m\sum_{\alpha_1,\cdots, \alpha_{N}} L_{[1]}^{\alpha_1} \cdots \left(C_{[\text{oc}]}^{\alpha_{\text{oc}}}\right)_k \cdots R_{[N]}^{\alpha_{N}} \ket{\alpha_1 \cdots \alpha_{N}}\ket{k}.
\label{eq:mtdmrg_equation}
\end{equation}
The proper orthonormality of the states represented by the MTMPS is ensured by the condition in~\cref{fig:mtdmrg_diags}b.
 
Instead of variationally optimizing a single MPS to represent the ground state of~$\hat{H}$ as per the standard DMRG, the multi-target DMRG algorithm optimizes the MTMPS to represent the~$m$ lowest energy eigenstates of the Hamiltonian. 
The MTDMRG protocol is that of the standard DMRG but with modifications of steps ii) and iii). In step ii), instead of choosing the groundstate of~$\hat{H}^{\text{eff}}_{x,x+1}$ as the optimal two-site state for sites~$x,x+1$, the lowest~$m$ energy eigenstates are combined to form a multi-target two-site effective state~$\gamma_{x,x+1}^{\text{opt}}$:
\begin{equation}
    \gamma_{x,x+1}^{\text{opt}} = \sum_{k=1}^m \left(\psiopt\right)_k \ket{k}.
\label{eq:twosite_psieff_mt}
\end{equation}

In step iii), this object is subsequently decomposed and truncated using an SVD, see ~\cref{fig:mtdmrg_diags}c. Compared with the standard DMRG algorithm, the only difference is that the object being decomposed has an extra leg that must be combined with those on the right before performing the SVD (or with those on the left for a left sweep).

The center of orthogonality is moved over one site and the update is repeated in sweeps until the total energy of the combined states is converged. The resulting state after MTDMRG is an MTMPS of the lowest~$m$ eigenstates of~$\hat{H}$. Obtaining several eigenstates at once through the MTDMRG offers a more uniform and efficient convergence across all eigenstates, compared with the orthogonalization approach. However, the bond dimension~$\chi$ of the MTDMRG has to be large enough in order to represent all the multi-targeted states, which for some Hamiltonians can be a significant increase in computational cost.
\\

\textit{DMRG-X---}The main drawback of the standard DMRG and MTDMRG methods for finding excited states is that in order to target a particular state of interest we are forced to find all eigenstates with energy lower than our target state. This incurs unnecessarily high memory and runtime costs. An illustrative example of this tradeoff is shown in~\cref{app:vsMTs}. 

We can target individual eigenstates of interest using the recently developed DMRG-X algorithm~\cite{Khemani_2016}. This algorithm takes as input not only the Hamiltonian but also a reference state, and finds the eigenstate of~$H$ that has the highest overlap with the reference state.  This approach is particularly useful when the eigenstates of~$H$ are close to product states, but with a small amount of delocalization; using a reference product state (\textit{bare} state) as input, we efficiently find the \textit{dressed} version of the state that remains localized but is an eigenstate of the many-body~$H$, including small fluctuations. We note that DMRG-X can also work if an eigenstate of interest is known to have high overlap with a simple reference state that is not a product state.  In that case the final dressed state found by the algorithm may not be localized, but it can still be targeted by its overlap with the reference state.  In our circuit-QED simulations using both DMRG-X and our new MTDMRG-X algorithm, and in~\cref{eq:target_bare_states} below, we always use product states as the bare states, but generalizing is easy.

DMRG-X was originally proposed for studying MBL systems.  However, it is also useful for our circuit-QED setup.  In circuit QED, we expect the eigenstates that form the dressed computational basis to be localized~\cite{berke2022transmon}. This localized regime is also desired for quantum control. The localization of the eigenstates implies that they have high overlap with product bare states  of the system. 

In practice, the algorithm is the same as DMRG except with a small modification to step ii).  
The variational MPS is initialized to be the product (bare) state that has the largest overlap with our target state. The rule for finding the optimal two-site effective state~$\ket{\psi^{\text{opt}}}_{x, x+1}$ in step ii) becomes: find the eigenstate of~$\hat{H}^{\text{eff}}_{x, x+1}$ with largest overlap with the current variational MPS. 
This eigenstate can be found by computing low-energy eigenstates of~$\hat{H}^{\text{eff}}_{x, x+1}$ using Lanczos-type methods, iteratively adding further excited states until one is found that has the desired high overlap.

\section{\MakeUppercase{Multi-target DMRG-X algorithm}}
\label{sec:mtdmrgx}

 Our new algorithm for finding excited states extends the DMRG-X method by utilizing the multi-target MPS ansatz with a two-site update rule where the overlap with a set of bare states, rather than just a single reference state, is maximized. We call this approach multi-target DMRG-X (MTDMRG-X). A crucial advantage of this method with respect to the usual single-target DMRG-X is the efficient simultaneous resolution of strongly resonant states, where a set of wavefunctions have almost equal projection onto the same bare states (e.g.~$\psi_{\pm}\sim\ket{01}\pm\ket{10}$). Such hybridized states are relevant in circuit QED for investigating effects such as cross-talk or state-dependent avoided crossings (see ~\cref{app:xy_splitting} and our results in~\cref{sec:results_mtdmrgx}). Performing this calculation with the conventional DMRG-X involves running separate instances with a new effective Hamiltonian that projects out the previously found resonant states, which comes at an increased computational cost. Below we outline the main steps of the MTDMRG-X algorithm for targeting the set of~$m$ dressed states given a set~$S$ of reference (bare) states:
\begin{multline}
    S = \{\ket{b_1},\ket{b_2}, \cdots, \ket{b_m}\},\\ \text{where}\,\, \ket{b_k} = \bigotimes_{x=1}^N\ket{p_{k,x}}\,\,
    \text{and}\,\, p_{k,x} \in \{0,\dots, d-1\},
\label{eq:target_bare_states}
\end{multline} 
where~$N$ is the number of sites and~$d$ is the number of bosonic levels of each site. The MPS is initialized by combining the product states of~$S$ into an MTMPS of dimension~$m$, such that:
    \begin{equation}
        \Gamma = \sum_{k=1}^{m}\ket{b_k}\ket{k}.
    \end{equation}

Following the standard DMRG procedure, we sweep back and forth, locally updating tensors.  For each pair of sites~$x,x+1$ on a right sweep, in step i) of the local update we obtain the effective Hamiltonian, $\hat{H}^{\text{eff}}_{x,x+1}$, exactly as in the standard DMRG algorithm. 

The core of the new algorithm lies in the modification of step ii). We have a set of reference bare states~$S$ (\cref{eq:target_bare_states}) which we project onto the MPS to obtain a set of bare state projections~$\{\ket{P_k}\}$. We would like to match each bare state to an eigenstate of the effective Hamiltonian. The effective states and bare states are matched by largest overlap. We may do this in several ways.  The standard approach is to sequentially explore the spectrum of~$\hat{H}^{\text{eff}}_{x,x+1}$ in ascending eigenvalue order, matching eigenstates to bare states as we progress. Alternatively, we can target the matching eigenstates more efficiently by building the Krylov subspace around the bare states during the Lanczos algorithm, an approach which we refer to as Lanczos-X. The computational runtime using this new Lanczos variant has weaker dependence on the number of excitations than other standard methods. See ~\cref{app:mtdmrg-x} for a more detailed description of MTDMRG-X optimal state finding and Lanczos-X. Once all the effective two-site states matching the bare states have been found, we construct the two-site multi-target state:
    \begin{align}
        \gamma^{\text{opt}}_{x,x+1} = \sum_{k=1}^m \left(\ket{\psi^{\text{opt}}}_{x, x+1}\right)_k\ket{k}\quad \text{where}\\
        \left(\ket{\psi^{\text{opt}}}_{x, x+1}\right)_k = \underset{\lambda \,\in \,\chi^2d^2}{\arg\max} \left| \braket{P_k \mid \psi^{(\lambda)}}_{x,x+1} \right|
    \label{eq:twosite_psieff_mtdmrgx}
    \end{align}

As in step iii) of the MTDMRG protocol, $\gamma^{\text{opt}}_{x,x+1}$ is decomposed and truncated as in ~\cref{fig:mtdmrg_diags}c. The variational MTDMRG-X algorithm continues until the total energy of the MTMPS has converged.\par 

In practice the full set of reference states~$S$ for a strongly hybridized state might not be known. We can solve this issue by running DMRG-X using as a reference one single bare state that we know is a strong support of the hybridized wavefunction. We then compute the overlap of the converged MPS with all product basis states within an energy range and find the full set of bare state support~$S$. This is a set of bare states such that the total overlap with~$\ket{\Psi}$ is above a threshold~$\theta$, meaning that we have found most of the basis state support of the hybridized set of states. Explicitly:
\begin{equation}
    S = \{ \ket{b_k}\} \quad s.t. \quad  \sum_{k=1}^m \left| \bra{b_k} \Psi \rangle \right|^ 2 > \theta
\end{equation}
We then run MTDMRG-X as described above using the set of states~$S$ to compute the full set of nearly-resonant eigenstates.

\section{\MakeUppercase{Modeling circuit-QED systems}}
\label{sec:circuit QED}

\subsection{Simplified device Hamiltonian}

Circuit quantum electrodynamics (circuit QED) studies the interaction of nonlinear superconducting circuit elements with quantized electromagnetic fields in the microwave regime~\cite{blais2004cavity,blais2021circuit}. In these circuits, the Josephson junction is the source of nonlinearity, leading to the formation of an anharmonic energy-level structure. Such a device can be used as a qubit by encoding information into the ground state~$\ket{0}$ and first excited state~$\ket{1}$, and selectively addressing the 1-0 transition for control. However, many other encodings and qubit-operation modalities are possible in the superconducting platform---see, e.g., Ref.~\cite{gyenis2021moving}.

In this work, we consider a circuit-QED architecture based on the flux-tunable transmon circuit~\cite{koch_2007}, consisting of a capacitively shunted dc-SQUID. The circuit parameters are the capacitive energy, $E_\mathrm{C}$, and the flux-tunable Josephson energy~$E_\mathrm{J}(\Phi_\mathrm{ext})$, where~$\Phi_\mathrm{ext}$ is the magnetic flux threaded by the dc-SQUID~\cite{krantz2019quantum}. The transmon regime is realized for~$E_\mathrm{J}(\Phi_\mathrm{ext})/E_\mathrm{C}\gg 1$, and is well described by the weakly anharmonic oscillator Hamiltonian (see~\cref{app:transmon_H}):
\begin{equation}
    \hat H_i/\hbar = \omega_i\hat a_i^\dagger \hat a_i -\frac{\eta_i}{2}\hat a_i^{\dagger 2} \hat a_i^2,
    \label{eq:qubit_hamiltonian}
\end{equation}
where~$\omega_i\approx \sqrt{8 {E_{\mathrm{C},i}}E_{\mathrm{J},i}(\Phi_{\mathrm{ext},i})}/\hbar-\eta_i$ is the mode frequency, $\eta_i\approx  E_{\mathrm{C},i}/\hbar$ is the anharmonicity, and~$(\hat a_i,\hat a_i^\dagger)$ is a set of bosonic ladder operators satisfying~$[\hat a_i,\hat a_i^\dagger]=1$. The subindex~$i\in[1,N]$ represents a mode indexing.

Between any two modes~$i$ and~$j$ in the circuit, the coupling is capacitive and is described by the Hamiltonian
\begin{equation}
    \hat H_{ij}/\hbar = g_{ij} (\hat a_i + \hat a_i^\dagger)(\hat a_j + \hat a_j^\dagger),
    \label{eq:coupling_hamiltonian}
\end{equation}
where~$g_{ij}$ is a coupling strength, typically expressed in terms of the coupling efficiency~$0\leq k_{ij}\ll 1$ and the respective mode frequencies as
\begin{equation}
    2g_{ij} = {k_{ij}}\sqrt{\omega_i \omega_j}.
    \label{eq:coupling_strength}
\end{equation}
The architecture of interest is comprised of qubits and couplers of transmon modality~\cite{neill2017path,Yan_2018}, and the tunable-coupling scheme is described in~\cref{app:tunable_coupler}. Including both qubit and coupler transmons, the total device Hamiltonian is
\begin{equation}
    \hat H = \sum_{i=1}^N \hat H_i + \sum_{\{i,j\}} \hat H_{ij},
    \label{eq:device_hamiltonian}
\end{equation}
with single-mode contributions~$\hat{H}_i$ and two-mode contributions~$\hat{H}_{ij}$ given by~\cref{eq:qubit_hamiltonian} and~\cref{eq:coupling_hamiltonian}, respectively.  The pairs~$\{i,j\}$ that contribute are determined by the chip connectivity. The eigenvalues and eigenvectors of~\cref{eq:device_hamiltonian} are referred to as \textit{dressed energies} and \textit{dressed states}, respectively. This is in contrast to the \textit{bare states}, which are tensor products of the uncoupled transmon states that separately diagonalize~\cref{eq:qubit_hamiltonian} on each site---i.e., the single-site Fock states. 

\begin{figure}[t!]
\centering
\includegraphics[width=1\columnwidth]{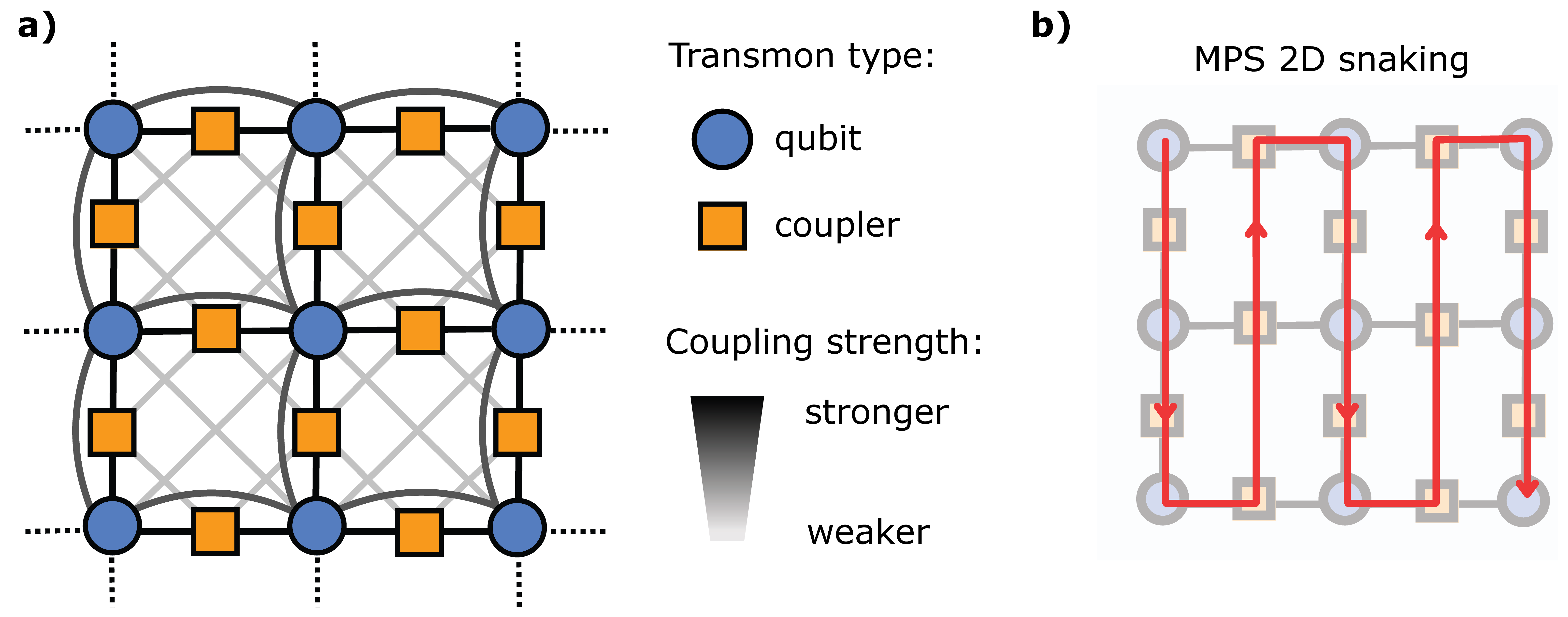}
\caption{\textbf{Transmon qubit and coupler lattice.} \textbf{\textsf{a)}} Two-dimensional chip cut-out showing the connectivity of the transmon system of interest. Black and dark-gray lines (vertical and horizontal) represent intended capacitive couplings between adjacent qubits and couplers. Light-gray lines (diagonal) represent stray couplings. The transparency of each line determines, qualitatively, the magnitude of the respective coupling strength. \textbf{\textsf{b)}} Example MPS mapping of a 3$\times$3 qubit array. The red path with arrows shows the site ordering in the 1D MPS.}
\label{fig:2d_chip_connectivity} 
\end{figure}

\subsection{Problem setup}
The computational basis states of the transmon system are defined to be a subset of the dressed states and, ideally, are sufficiently localized to enable high-fidelity quantum control via local transmon voltage and/or current biases. 

Having access to properties of the system's excitations is useful.  For instance, knowing the degree of delocalization helps us understand crosstalk error. Dressed frequencies yield information regarding coupling strengths (both intended and spurious) and state-dependent qubit-energy corrections (which are also a source of error)~\cite{berke2022transmon}.
Solving for the eigenvalues and eigenvectors of~\cref{eq:device_hamiltonian} using full state vectors is exponentially hard in the number of transmons, $N$. More precisely, the exact wavefunction has dimension~$\textstyle\prod_{x=1}^Nd_i$, where~$d_i$ is the local dimension (number of levels) considered for each transmon. Typically, $d>3$ for a realistic representation of the transmon circuit. Full-wavefunction, matrix-free methods leveraging the local structure of~\cref{eq:device_hamiltonian} can handle system sizes of order~$N\sim 20$ (for~$d=4$) in a reasonable time and using reasonable computational resources. This number of qubits is small in comparison to that of state-of-the-art devices, see e.g., Ref.~\cite{ai2024quantum}.

In this work, we develop a tensor-network approach to solve for the low-lying excitations of order-$100$ transmon Hamiltonians in 2D, pushing our results well beyond the exact-diagonalization regime. As argued in previous sections, our numerical technique leverages the quasi-local structure of the Hamiltonian eigenstates for efficient state targeting, instead of relying only on energy eigenvalues for that purpose. 

\subsection{Numerical results}
Here we present our numerical results obtained for a two-dimensional transmon array including qubits and couplers. The system Hamiltonian is specified in~\cref{eq:device_hamiltonian} with the connectivity shown in~\cref{fig:2d_chip_connectivity}a. The MPS and MPO encodings are done by ``snaking over'' the 2D array as shown in~\cref{fig:2d_chip_connectivity}b.

\subsubsection{Single-qubit and single-coupler excitations via DMRG-X}\label{sec:qubit_coupler_loc}
\begin{figure*}[t!]
\centering
\includegraphics[width=1\textwidth]{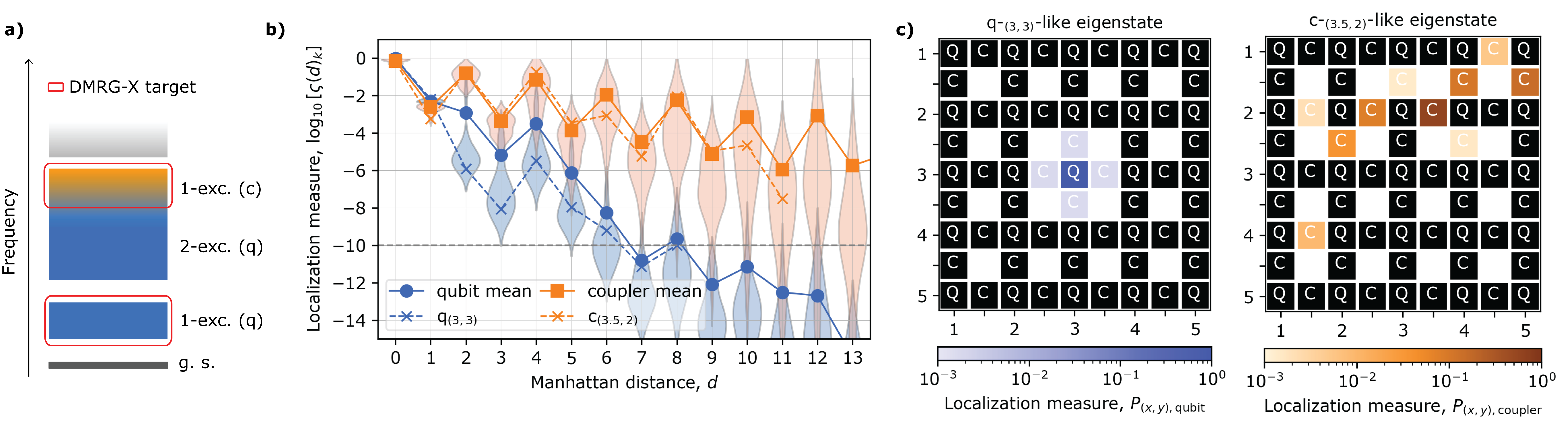}
\caption{\textbf{Localization measures in a 2D transmon array.} \textbf{\textsf{a)}} Sketch of the low-frequency sector of the device spectrum, which is structured in bands. The first band above the ground state corresponds to single-qubit excitations, while the second band is formed by two-qubit and (level-2) single-qubit excitations, followed by single-coupler excitations. We target single-qubit and single-coupler excitations using DMRG-X (red boxes), without the need to first compute eigenstates in between these two bands. \textbf{\textsf{b)}} Mean qubit (solid line, blue) and coupler (solid line, orange) localization measure as a function of Manhattan distance, $d$. Dashed lines show the localization metric (\cref{eq:combined_loc}) for two example cases of two eigenstates corresponding to the qubit located at~$(x,y)=(3, 3)$ (q$_{(3, 3)}$, in blue) and the coupler located at~$(x,y)=(3.5, 2)$ (c$_{(3.5, 2)}$, in orange). (The dashed gray line represents an approximate bound consistent with the requested DMRG accuracy; results below this bound are only shown to display the data's trend.) \textbf{\textsf{c)}} Qubit q$_{(3, 3)}$ and coupler c$_{(3.5, 2)}$ eigenstate projections onto bare single-excitation states belonging to each~$(x,y)$ site in the chip. Values below~$10^{-3}$ are shown in black. We observe that coupler excitations can have significant support on far away sites, indicating a nonnegligible degree of coupler-mode delocalization.}
\label{fig:all_localization_experiment}
\end{figure*}

Determining how the bare qubit and coupler excitations are dressed in the presence of capacitive couplings is crucial for identifying sources of error such as crosstalk and population transport due to effective (mediated) nonlocal couplings. While a localized regime is desired for high-fidelity quantum control, strong two-qubit interactions---which in our architecture are largely due to the couplers---are required for realizing fast two-qubit gates. Naturally, these engineered interactions can introduce some degree of mode delocalization. We study this effect numerically in a~$5\times5$ chip, including 25 transmon qubits and 40 transmon couplers. 

In our simulation, the coupler frequency between two adjacent qubits is adjusted to turn `off' the effective exchange coupling between the qubits when they are brought into resonance, see~\cref{app:xy_splitting} and~\cref{app:tunable_coupler}. This is done for each qubit pair and its respective coupler, without accounting for the rest of the modes in the circuit when doing so. The bare qubit frequencies are set before adjusting the coupler frequencies, and selected in the range~$6-6.6\,$GHz (see~\cref{fig:loc_exp_qfrequencies} for details). While these frequencies values are typical in circuit QED, we have sampled them randomly for our simulations and they do not correspond to a real experimental frequency configuration. Setting the coupler frequencies to their respective `off' values defines the `idle' operating regime where qubit interactions are minimized.\\

\textit{Numerical benchmark---}We use DMRG-X to target single-excitation eigenstates of the qubit and coupler modes independently (this can be done in parallel), as shown in~\cref{fig:all_localization_experiment}a. 
For each mode, the reference state used to initialize the DMRG-X simulation is the product state where the targeted transmon is in the state~$\ket{1}$ and all others are in~$\ket{0}$. 
Convergence of DMRG-X is assessed using the effective state energy difference across consecutive sweeps, using the convergence threshold~$10^{-10}$ GHz (see~\cref{app:convergence_criteria}). For a bond dimension~$\chi = 80$ the metric~\cref{eq:psi_hamiltonian_variance} yields Hamiltonian variances in the range~$10^{-9} < \text{var}(\hat{H}/h)<  10^{-7}$ (GHz$^2$). See~\cref{app:numerical_details} for further experimental details.

We study eigenstate localization via the intuitive per-site localization measure:
\begin{equation}
    P_{(x,y),k} = |\braket{0,\dots, 1_{(x,y)},\dots, 0|\Psi_k}|^2,    
    \label{eq:persite_loc}
\end{equation}
where~$|0\dots, 1_{(x,y)},\dots, 0\rangle$ represents a bare state with a single excitation in site~$(x,y)$, and~$|\Psi_k\rangle$ is an eigenstate of interest, labeled by~$k$. We then aggregate the values of~$P_{(x,y),k}$ into the combined quantity
\begin{align}
    \varsigma(d)_k = \sum_{(x,y) \in B_d} P_{(x,y),k},
    \label{eq:combined_loc}
\end{align}
where~$B_d = \{(x,y) | \text{Man}[(x,y),(x_k,y_k)] = d\}$ is the set of points located at fixed Manhattan distance~$d$ from~$(x_k,y_k)$. Here, $(x_k,y_k)$ is the position we associate with eigenstate~$k$, corresponding to the lattice site where~$|\Psi_k\rangle$ has maximum support. We distinguish between~\cref{eq:combined_loc} for qubit-like [$\varsigma(d)_{k,\mathrm{q}}$] and coupler-like [$\varsigma(d)_{k,\mathrm{c}}$] eigenstates, for which~$(x_k,y_k)$ is a qubit or coupler site, respectively, because the two cases exhibit very different localization profiles. 

We plot the aggregated localization metrics in~\cref{fig:all_localization_experiment}b for the qubits (blue) and couplers (orange). Solid lines show the mean values of~$\varsigma(d)_{k,\mathrm{q}}$ and~$\varsigma(d)_{k,\mathrm{c}}$ as a function of Manhattan distance, $d$. We also show the resulting distribution for all qubits and couplers (via the so-called ``violin" plots), and the Manhattan distance aggregated localization plot for two representative qubit and coupler examples: qubit~q$_{(3,3)}$ and coupler~c$_{(3.5,2)}$ (dashed lines). Note that what is plotted is the base-10 logarithm of the localization measure, so that the full range and trend of the decay of excitation probability with distance can be seen clearly.

In addition, \cref{fig:all_localization_experiment}c shows the per-site localization measure in~\cref{eq:persite_loc} for the eigenstates associated with a single excitation on~q$_{(3,3)}$ and on~c$_{(3.5,2)}$. While the qubit-like eigenstate is largely localized, we observe nonzero support of the coupler-like eigenstate away from site~$(3.5,2)$ in the grid. (This behavior is consistent with the aggregated metric~$\varsigma(d)_{k,\mathrm{c}}$, shown in~\cref{fig:all_localization_experiment}b.) 

Understanding the origin of coupler-mode delocalization for our parameter set of choice is beyond the scope of this work. However, the effect is likely a result of the relative proximity of coupler frequencies, combined with spurious direct (see~\cref{fig:2d_chip_connectivity}a) and indirect (qubit-mediated) coupler-coupler couplings. The weak detuning between coupler frequencies is a consequence of the strategy we use to set them: given nearly uniform qubit-coupler and qubit-qubit couplings, we determine, for all pairs of adjacent qubits, the coupler frequencies that turn `off' the resonant qubit-qubit coupling. This leads to coupler frequencies that are weakly detuned from each other. As a consequence, any stray and qubit-mediated coupler-coupler coupling can lead to delocalization of the coupler excitations. We stress that the coupler-mode delocalization we observe is highly dependent on the frequency and coupling-strength parameters that we selected and does not reflect an experimental device. 

\subsubsection{State-dependent two-qubit couplings via MTDMRG-X} \label{sec:results_mtdmrgx}

\begin{figure*}[t!]
\centering
\includegraphics[width=1\textwidth]{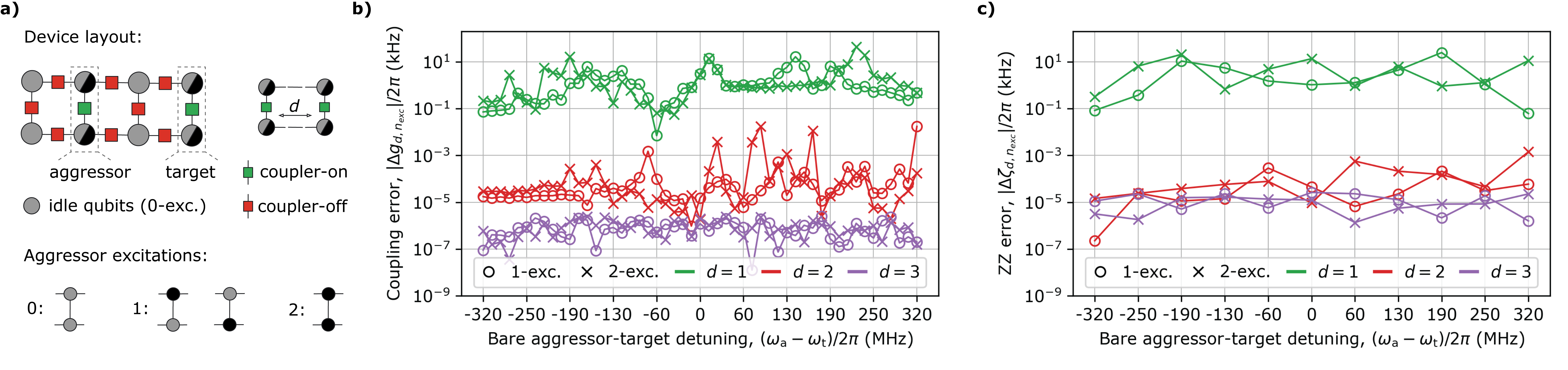}
\caption{\textbf{State-dependent single- and double-excitation couplings on a target qubit pair.} \textbf{\textsf{a)}} Sketch of the numerical experiment: the couplings ($g$ and~$\zeta$) between the two qubits in a target pair of interest are monitored as a function of the distance to and frequency and number of excitations of an aggressor qubit pair. The rest of the qubits and couplers are at idle, with no excitations. \textbf{\textsf{b)}} Aggressor-state-dependent target-pair exchange coupling~$g$ as a function of aggressor-target detuning, $\omega_\mathrm{a}-\omega_\mathrm{t}$, for a set of distances between the target and aggressor pairs. We show the absolute value of the difference between the coupling calculated for no excitations in the aggressor pair (reference value) versus the value extracted for one (aggressor states close to~$\ket{01}$ and~$\ket{10}$) and two (aggressor state close to~$\ket{11}$) excitations. The state-dependence of these quantities demonstrates the presence of weak longer-ranged effective couplings. \textbf{\textsf{c)}} Same as~\textbf{\textsf{b)}}, but for the ZZ coupling, $\zeta$.}
\label{fig:all_mtdmrgx_experiment}
\end{figure*}

State localization is an instructive metric, but it only provides a qualitative picture of potential error channels. A more direct probe is offered by the single- and double-excitation state couplings, which in circuit QED are typically referred to as exchange and ZZ (or cross-Kerr) interactions, and denoted by~$g$ and~$\zeta$, respectively.

The single-excitation exchange coupling between two modes~$k$ and~$l$ is defined as
\begin{equation}
    2g =  \underset{\omega^\mathrm{b}_{k}}{\arg\min} \left[|\omega_{k}-\omega_{l}|\right],
    \label{eq:g_def}
\end{equation}
where~$\omega^\mathrm{b}_{k}$ is the bare frequency of qubit~$k$, and~$\hbar\omega_{k}$ and~$\hbar\omega_{l}$ are the energy eigenvalues corresponding to the single-excitation states localized in each mode, see also~\cref{app:xy_splitting}. While obtaining direct estimates of~$g$ is possible via block-diagonalization strategies (e.g., Schrieffer-Wolff theory~\cite{magesan2020effective}), \cref{eq:g_def} can also be regarded as a numerical procedure: sweep the bare frequency~$\omega^\mathrm{b}_{k}$ while recording the dressed frequency difference~$\omega_{k}-\omega_{l}$ (detuning). Provided the qubit-$k$-like and qubit-$l$-like eigenvalues have an avoided-level crossing during this sweep, the minimum detuning (resonance condition) is a measure of the effective two-qubit coupling~$g$, as defined in~\cref{eq:g_def}.

In practice, the coupling between two qubits~$(k,l)$ sharing a coupler is measured as a function of the coupler frequency. The coupler frequency is a bias parameter that can both turn `off'~$(g=0)$ and `on' ($g\neq 0$) the qubit-qubit interaction to enable fast two-qubit gates with high on-off ratio (see ~\cref{app:tunable_coupler}). A common procedure for calibrating~$g$ between a pair of qubits of interest considers the other qubits in the processor---which we refer to as the `environment' or `spectators'~\cite{krinner2020benchmarking}---in their ground state. The calibrated value of~$g$ is used to determine, for example, pulse sequences for two-qubit gates.  However, during device operation, spectator qubits can be in any state, including being entangled with the pair of interest. A dependence of~$g$ on the environment state results in the calibrated pulse sequence no longer accurately implementing the intended gate, introducing a coherent-error channel. 
Such in-context error mechanisms can significantly degrade the fidelity of quantum algorithms with respect to what is expected from isolated single- and two-qubit gate fidelities.

A complementary metric to the exchange interaction is the ZZ coupling between two qubits~$k$ and~$l$, defined as
\begin{equation}
    \zeta = (\omega_{1_k 1_l} - \omega_{0_k 1_l}) - (\omega_{1_k 0_l} - \omega_{0_k 0_l}).
    \label{eq:zeta_def}
\end{equation}
In other words, $\zeta$ quantifies how much the transition frequency of qubit~$k$ depends on the state of qubit~$l$. While a nonzero ZZ interaction can be used for two-qubit gates, a nonzero~$\zeta$ value at idle represents a coherent error. 

Similarly to the exchange coupling, the ZZ interaction can depend on the state of spectator qubits surrounding the qubit pair of interest. Such state dependence hints at correlated errors of higher weight, which should also be minimized for high-fidelity quantum algorithms. \\

\textit{Numerical benchmark---}In~\cref{fig:all_mtdmrgx_experiment}, we investigate the dependence of the exchange and ZZ couplings between a pair of adjacent qubits (the \emph{target} pair) with respect to the state of two nearby qubits on the chip (the \emph{aggressor} pair).  The qubits in the aggressor pair are on resonance with each other, and the qubits in the target pair are at a relative detuning~$\delta$ that produces the minimum avoided crossing between them, mimicking the calibration procedure associated with~\cref{eq:g_def}. The coupler within each pair is turned on, emulating simultaneous gates occurring on the two pairs.  The remaining qubits in the system, the spectators, are off-resonance and have no excitations, and all couplings outside of the target and aggressor pairs are turned off. This setup is shown in~\cref{fig:all_mtdmrgx_experiment}a and further numerical details are provided in~\cref{app:numerical_details}.

\Cref{fig:all_mtdmrgx_experiment}b shows the correction to the exchange coupling between the qubits in the target pair as a function of the detuning with respect to the aggressor pair. The result is shown for separation distances 1, 2, and 3 between the target and aggressor pairs, and as a function of the number of excitations in the aggressor pair: 1 (states~$\ket{01}$ and~$\ket{10}$ of the aggressor pair, approximately) and 2 (state~$\ket{11}$, approximately). Note that the aggressor-pair frequency crosses the target-pair frequency, leading to a subspace with up to four states that can become resonant. Therefore, targeting of the subspaces in~\cref{table:targeted_subspaces} ($g$ estimation row) is required.  The computation is done efficiently with the MTDMRG-X algorithm.

Similarly, \cref{fig:all_mtdmrgx_experiment}c shows the state-dependence of the ZZ interaction between the qubits of the target pair, as a function of the detuning from the aggressor pair, distance, and number of excitations. In this case, the targeted subspace of states is provided in~\cref{table:targeted_subspaces} ($\zeta$ estimation row). 

For the selected parameter set, the state-dependence of both exchange and ZZ couplings is nonzero, but the values we arrive at are in this case mostly too small to be a dominant source of error. An exception is the ZZ coupling at distance~$d=1$, where corrections due to the aggressor pair can reach~10 MHz when considering two excitations. However, operating two-qubit gates at such close proximity is often avoided in quantum algorithms. Note also that the values we report are both a function of the parameter set and the device layout we considered. In the absence of couplers, longer range interactions can become more problematic~\cite{berke2022transmon}. 

\begin{table}[]
\centering
\begin{tabular}{c|c|c|c|}
 &~$n_\mathrm{exc}=0$ &~$n_\mathrm{exc}=1$ &~$n_\mathrm{exc}=2$ \\
\hline

$g$ estimation &~$\begin{Bmatrix}
        (0,0,1,0)\\
        (0,0,0,1)
      \end{Bmatrix}$ &~$\begin{Bmatrix}
                        (1,0,1,0)\\
                        (0,1,1,0)\\
                        (1,0,0,1)\\
                        (0,1,0,1)
                      \end{Bmatrix}$ &~$\begin{Bmatrix}
                                        (1,1,1,0)\\
                                        (1,1,0,1)
                                     \end{Bmatrix}$ \\
\hline
$\zeta$ estimation &~$\begin{matrix}
            (0,0,0,0)\\
            
            \begin{Bmatrix}
            (0,0,1,0)\\
            (0,0,0,1)
          \end{Bmatrix}\\
          (0,0,1,1)
          \end{matrix}$     
          &$
          \begin{matrix}
          \vspace{2mm}
          \begin{Bmatrix}
              (1,0,0,0)\\
            (0,1,0,0)
          \end{Bmatrix}\\
          \vspace{2mm}
          \begin{Bmatrix}
            (1,0,1,0)\\
            (0,1,1,0)\\
            (1,0,0,1)\\
            (0,1,0,1)\\
         \end{Bmatrix}\\
         \begin{Bmatrix}
              (1,0,1,1)\\
            (0,1,1,1)
          \end{Bmatrix}
         \end{matrix}$ 
         
         & ~$\vspace{2mm} \begin{matrix}
            (1,1,0,0)\\
             \begin{Bmatrix}
            (1,1,1,0)\\
            (1,1,0,1)
         \end{Bmatrix}\\
         (1,1,1,1)
         \end{matrix}$\\
         
\hline
\end{tabular}

\caption{Targeted (bare) state subspaces used for the exchange ($g$) and ZZ ($\zeta$) coupling estimation in~\cref{fig:all_mtdmrgx_experiment}, as a function of the number of excitations ($n_\mathrm{exc}$) in the aggressor-qubit pair. State indexing: (aggressor-top, aggressor-bottom, target-top, target-bottom). Bracketed sets of bare states are targeted simultaneously using MTDMRG-X given the strong resonance regime. Qubits and couplers not explicitly included in the state indexing that is shown are assumed to be in the ground state.}
\label{table:targeted_subspaces}
\end{table}  

\section{\MakeUppercase{Conclusions}}
\label{sec:conclusions}

In this work, we have introduced a new density-matrix-renormalization-group algorithm to simultaneously target multiple states via a state-overlap-based tensor-update objective function. This method, which we call MTDMRG-X, is an extension of two algorithms previously known in the literature, namely MTDMRG and DMRG-X. Crucially, MTDMRG-X allows for the targeting of highly excited states in a many-body system without the need to first compute lower-energy states and in cases where excited states are strongly hybridized, leading to the formation of a target subspace. Moreover, by modifying the Lanczos algorithm, we target wavefunctions deep into the energy spectrum with a significantly reduced runtime in comparison to the standard Lanczos approach.

Our numerical method is motivated by its direct application to the modeling of circuit-quantum-electrodynamics systems. In particular, we demonstrate how state-overlap-based objective functions in DMRG-X are useful to resolve single-excitation eigenstates centered on qubits and couplers in a transmon device. Moreover, we show that while qubit excitations are largely localized, couplers---which nominally remain in their ground state---can exhibit delocalized excitations. Finally, we demonstrate how combining state-overlap- and multi-state-targeting in MTDMRG-X enables the study of state-dependent exchange and ZZ couplings in circuit QED, even in situations where multiple states become resonant. Note that our conclusions apply to a typical parameter set selected for simulations and are not necessarily representative of the behavior of an experimental device. 

We expect MTDMRG-X to become a useful tool for studying many-body systems and, in particular, enabling more effective design of large-scale circuit-QED devices.

\begin{acknowledgments}
The authors thank B. Kobrin, W. P. Livingston, E. Rui, A. T. Petrescu, N. Astrakhantsev, Y. D. Lensky, A. Klots, M. I. Dykman, and V. N. Smelyanskiy for insightful discussions. S.G.G is partially supported by Fundaci\'on Rafael del Pino. G.V. is a CIFAR associate fellow in the Quantum Information Science Program, a Distinguished Invited Professor at the Institute of Photonic Sciences (ICFO), and a Distinguished Visiting Research Chair at Perimeter Institute.

\textit{Note--}During the completion of this work, we became aware of a similar work with applications to the fluxonium qubit~\cite{Rui_inprep}.

\end{acknowledgments}

\bibliography{main}
\appendix

\section{\MakeUppercase{Multi-target DMRG-X algorithm}}
\label{app:mtdmrg-x}

As introduced in the main text, the MTDMRG-X algorithm combines the multi-target MPS ansatz of the MTDMRG protocol with the spatial update rule of DMRG-X, allowing the simultaneous successful resolution of multiple potentially hybridized eigenstates. Here we provide more details about the update rule of MTDMRG-X, regarding how the two-site effective multi-target state is obtained. Recall that this is just a modification of step ii) in the standard DMRG algorithm (\cref{point:choice_heff}).\\
Consider an MTDMRG-X update on sites~$x,x+1$. Given a set of~$m$ reference bare states~$S=\{\ket{b_k}\}$\footnote{We note again that the bare states need not be product states.} we first define the set of projections~$\{\ket{P_k}\}$ of the current variational MPS onto the bare states. The computation of the projections is depicted in ~\cref{fig:mtdmrgx_update_diagram}a. We then construct the two-site effective Hamiltonian~$\hat{H}^{\text{eff}}_{x,x+1}$, as in step i) of the standard DMRG in the main text. The core of the MTDMRG-X algorithm lies in the way we obtain the optimal two-site eigenstates of~$\hat{H}^{\text{eff}}_{x,x+1}$ that make up the optimal two-site multi-target state~$\gamma_{x,x+1}^{\text{opt}}$. These are the~$m$ eigenstates of the effective Hamiltonian with largest overlap with the bare state projections~$\{\ket{P_k}\}$.\\

In the original DMRG-X paper~\cite{Khemani_2016}, the effective Hamiltonian~$\hat{H}^{\text{eff}}_{x, x+1}$ is exactly diagonalized (with cost~$\mathcal{O}(\chi^6)$), and the optimal two-site effective state~$\ket{\psi^{\text{opt}}}_{x,x+1}$ is obtained by computing all eigenstate overlaps with the reference state and selecting the largest. There can be more efficient ways of finding the set of highest-overlap states~$\gamma_{x,x+1}^{\text{opt}}$ depending on the MPS bond dimension and how deep into the spectrum of~$\hat{H}^{\text{eff}}_{x,x+1}$ the optimal states are located. Here we will describe two Lanczos methods that can be used for finding the states. The first one is a sequential Lanczos approach, where the spectrum of~$\hat{H}^{\text{eff}}_{x, x+1}$ is explored in ascending eigenvalue order using the usual Lanczos method. This is a convenient approach if the targeted eigenstates are low in the spectrum of~$\hat{H}^{\text{eff}}_{x, x+1}$. The second method we introduce builds the Krylov subspace in the Lanczos method around the bare states. This approach, which we refer to as Lanczos-X, offers a significant time advantage when targeting states with a large number of excitations, namely those deep in the spectrum of~$\hat{H}^{\text{eff}}_{x, x+1}$.

\subsection{Sequential Lanczos}

\begin{figure}[t!]
\centering
\includegraphics[width=1\columnwidth]{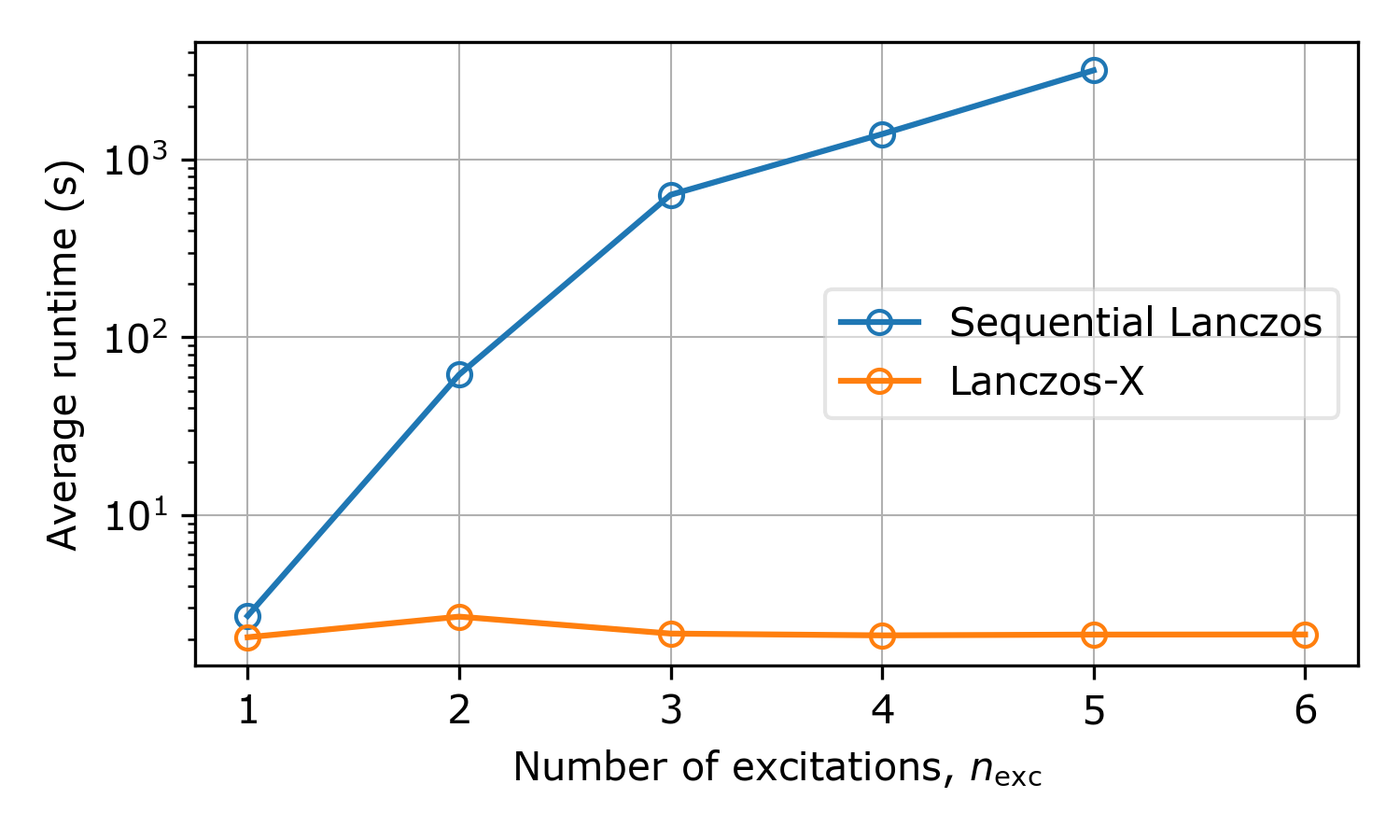}
\caption{Benchmark of the average time taken to run DMRG-X until convergence using sequential Lanczos (blue) and Lanczos-X (orange) to obtain the optimal two-site state. We show the runtime as a function of number of excitations in the target state. The average is taken over 14 target states for each excitation number. The system consists of a chain of 27 transmons (14 qubits, 13 couplers) each with local dimension 4.}
\label{fig:lanczosx_vs_sequential} 
\end{figure}
We find the eigenstates of~$\hat{H}^{\text{eff}}_{x,x+1}$ sequentially in ascending eigenvalue order, using the Lanczos algorithm~\cite{lanczos_1950}. In short, the Lanczos algorithm is an iterative method used to find the~$l$ `extreme' (highest or lowest) eigenvalues and associated eigenvectors of a Hermitian square matrix~$A$. In our case~$A = \hat{H}^{\text{eff}}_{x,x+1}$. The outline of the Lanczos method is as follows:
\begin{itemize}
    \item[\textit{i)}] Consider a normalized random initial seed state~$\ket{\psi}_{x,x+1}$.
    \item[\textit{ii)}] Build the Krylov subspace of dimension~$D$:
    \begin{align}
        V_{D} = \text{span}\big\{ &\ket{\psi}_{x,x+1}, \hat{H}^{\text{eff}}_{x,x+1}\ket{\psi}_{x,x+1}, (\hat{H}^{\text{eff}}_{x,x+1})^2\ket{\psi}_{x,x+1}, \nonumber \\ &\dots, (\hat{H}^{\text{eff}}_{x,x+1})^{D-1}\ket{\psi}_{x,x+1} \big\}
    \end{align}
    \item[\textit{iii)}] Diagonalize~$\hat{H}^{\text{eff}}_{x,x+1}$ projected onto~$V_D$.
    \item[\textit{iv)}] Increase~$D$ until the~$l$ extreme (lowest energy) eigenvalues have converged.
\end{itemize}
The Lanczos algorithm therefore constructs an orthonormal basis for the Krylov subspace. When using this method to find the eigenstates of ~$\hat{H}^{\text{eff}}_{x,x+1}$ in ascending eigenvalue order, we can set the number of Lanczos-resolved eigenstates to be~$l\geq 1$. The exploration is most efficient the closer this value is to the position of the two-site optimal states~$\ket{\psi_{x,x+1}^{\text{opt}}}$ in the spectrum of~$\hat{H}^{\text{eff}}_{x,x+1}$. 

\begin{figure*}[t!]
\centering
\includegraphics[width=1\textwidth]{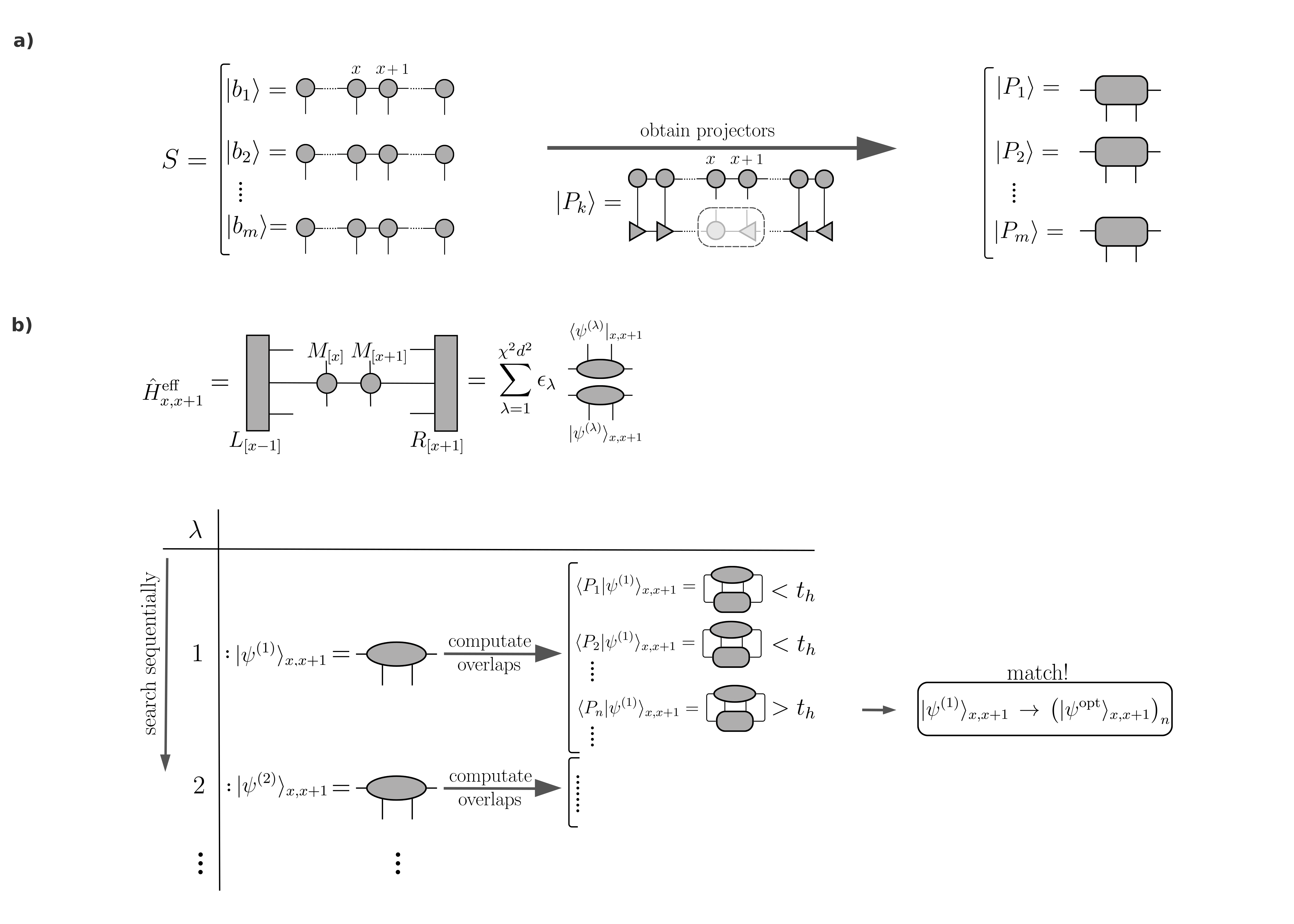}
\caption{\textbf{\textsf{a)}} Construction of the bare state projection~$\{\ket{P_k}\}$ for sites~$x,x+1$ of the MPS. \textbf{\textsf{b)}} Protocol for choosing the optimal two-site effective states using the sequential Lanczos algorithm for MTDMRG-X. The spectrum of~$\hat{H}^{\text{eff}}_{x, x+1}$ is explored sequentially and the overlap with all bare states is computed. In this example, the first eigenstate of the effective Hamiltonian, $\ket{\psi^{(1)}}_{x,x+1}$, is matched with bare state~$\ket{b_n}$.}
\label{fig:mtdmrgx_update_diagram} 
\end{figure*}

For concreteness, we explain in detail the MTDMRG-X algorithm using this standard Lanczos when setting~$l=1$, meaning we find one eigenstate of~$\hat{H}^{\text{eff}}_{x,x+1}$ at a time. As each eigenstate is found, we obtain its overlap with all product states in~$S$. If the overlap squared with a given bare state projection~$\ket{P_n}$ is above a certain threshold~$t_h$, that eigenstate is matched with a bare state and stored as ~$(\ket{\psi^{\text{opt}}}_{x, x+1})_n$. See ~\cref{fig:mtdmrgx_update_diagram}b. We continue to sequentially find eigenstates of~$\hat{H}^{\text{eff}}_{x,x+1}$ until either all the~$m$ states in~$S$ have been matched to a dressed state or a cumulative overlap with all states in~$S$ has been reached. In the latter case, the remaining unaccounted-for overlap for each unmatched state in~$S$ is less than the largest overlap between that state in~$S$ and a previously found eigenstate; that eigenstate is therefore matched with the state in~$S$. We then construct the two-site effective multi-target state:
\begin{equation}
        \gamma^{\text{opt}}_{x,x+1} = \sum_{k=1}^m \left(\ket{\psi^{\text{opt}}}_{x, x+1}\right)_k\ket{k}\quad.
\end{equation}

\subsection{Lanczos-X: restriction of Krylov subspace} 
\label{app:lanczos-x}
In the MTDMRG-X with standard (sequential) Lanczos as described above, we explore the effective Hamiltonian spectrum sequentially in ascending eigenvalue order until the state is matched or a cumulative overlap is saturated. Although this has a better scaling than the original exact diagonalization solution for low-lying excitations, it is still inefficient when the states we are targeting are deep in the spectrum, which occurs when the number of excitations in the targeted wavefunction is large.

A more efficient approach is to build the Krylov subspace in the Lanczos algorithm around the reference bare states in~$S$. We refer to this variant as \textit{Lanczos-X}. For example, for targeting one state using reference state~$\ket{b_k}$ we build the Krylov subspace around~$\ket{P_k}$ (projection of~$\ket{b_k}$ on the variational MPS):
\begin{align}
    V_D = \text{span}\big\{\ket{P_k}, (\hat{H}^{\text{eff}}_{x,x+1})\ket{P_k}, (\hat{H}^{\text{eff}}_{x,x+1})^2\ket{P_k}, \nonumber \\
    \dots, (\hat{H}^{\text{eff}}_{x,x+1})^{D-1}\ket{P_k}\big\}
\end{align} 
The optimal effective state~$\ket{\psi^{\text{opt}}}_{x, x+1}$ is approximated by the eigenstate of~$\hat{H}^{\text{eff}}_{ }$ that has the greatest overlap with~$\ket{P_k}$, and
the Krylov subspace dimension~$D$ is increased until~$\ket{\psi^{\text{opt}}}_{x, x+1}$ converges. This approach has a complexity of~$\mathcal{O}(\chi^3)$, dominated by the application of the effective Hamiltonian to the state. When targeting multiple states in MTDMRG-X, we can build the Krylov subspace as described in Ref.~\cite{baker2021}, Eq.~(21), then select the~$m$ eigenstates with the largest overlap with the reference state projections \{$\ket{P_k}$\}.

We benchmark the time required to find an eigenstate with~$n$ excitations using the original search method (ascending eigenvalue sequential search) and the adapted Lanczos-X for a 1D system of 25 transmons at idle, see~\cref{fig:lanczosx_vs_sequential}. For Lanczos-X, we used a Krylov subspace of dimension~$D=100$. 
We observe how the choice of the two-site update step can significantly speed up runtime. Indeed, we note a significantly improved scaling of computation time as a function of the number of excitations with Lanczos-X, apparently running in constant time in~$n$ in the explored regime.

\section{\MakeUppercase{DMRG convergence and accuracy}}
\label{app:convergence_criteria}

\subsection{Convergence criteria}
\label{app:evaluating_convergence}

The DMRG algorithm is assessed for convergence using the total energy of the effective two site state: 
\begin{equation}
\epsilon^{\text{opt}} = \bra{\psi^{\text{opt}}}_{x,x+1} \hat{H}^{\text{eff}}_{x,x+1} \ket{\psi^{\text{opt}}}_{x,x+1},
\end{equation}
for DMRG-X and 
\begin{equation}
   \sum_{k=1}^{m}\epsilon^{\text{opt}}_k \! = \!\sum_{k=1}^{m}\bra{k}\!\left(\!\bra{\psi^{\text{opt}}}_{x,x+1}\right)_k\hat{H}^{\text{eff}}_{x,x+1}\left(\ket{\psi^{\text{opt}}}_{x,x+1}\right)_k\!\ket{k}
\end{equation}
for MTDMRG and MTDMRG-X.  In particular, we test for convergence when the update site is in the middle of the chain, $x=N/2$. \\

\subsection{State accuracy}\label{app:mps_accuracy}
The resulting converged MPS state from the DMRG algorithm, $\ket{\Psi}$, will be a superposition of the actual eigenstate which we targeted~$\ket{\Tilde{\Psi}}$ (which has energy~$E_{\Tilde{\Psi}}$), and the rest of the energy eigenstates~$\{\ket{\phi_n}\}$:
\begin{equation}
    \ket{\Psi} = \sqrt{1-\varepsilon} \ket{\tilde{\Psi}} + \sqrt{\varepsilon}\sum_{n}c_n \ket{\phi_n} \quad \!\!\text{where}\!\!\quad \sum_{n}|c_n|^2=1 ,
\end{equation}
and where the index~$n$ runs over all eigenstates except for the index corresponding to~$\ket{\Tilde{\psi}}$. The infidelity of the state is therefore:
\begin{equation}
    1- |\braket{\Psi | \tilde{\Psi}}|^2 = \varepsilon.
\end{equation}
The error in the estimated energy eigenvalue is:
\begin{equation}
    \Delta E = \braket{\Psi| \hat{H} | \Psi} - E_{\tilde{\Psi}} = \varepsilon \sum_{n} \left| c_n \right|^2 \left( E_{n} - E_{\tilde{\Psi}}\right).
\label{eq:app_err_exp}
\end{equation}
The variance of the Hamiltonian, as introduced in ~\cref{eq:psi_hamiltonian_variance}, for~$\ket{\psi}$ can be written as:
\begin{equation}
    \text{var}(\hat{H},\ket{\Psi}) = \varepsilon \sum_{n} \left| c_n \right|^2 \left( E_{n} - E_{\Tilde{\Psi}}\right)^2.
\label{eq:app_var_exp}
\end{equation}
We can see from~\cref{eq:app_err_exp} and~\cref{eq:app_var_exp} that the variance and energy error have the same dependence on the fidelity. Therefore, the variance of the Hamiltonian gives us an idea of the magnitude of the energy error.

\section{\MakeUppercase{MTDMRG vs DMRG-X}}
\label{app:vsMTs}

\begin{figure}[t!]
\centering
\includegraphics[width=1\columnwidth]{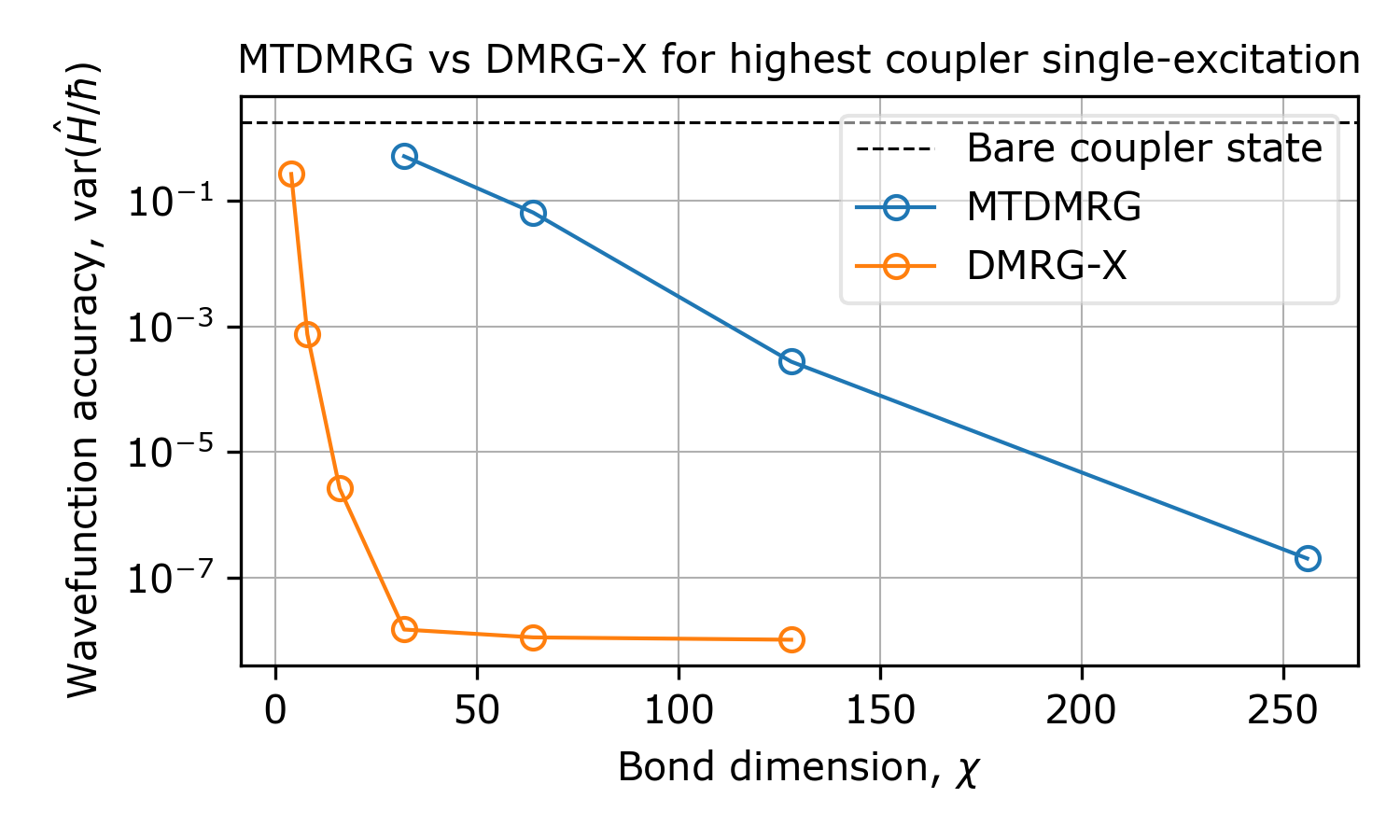}
\caption{Comparison of wavefunction accuracy, as measured by the variance of the Hamiltonian, for the highest coupler excitation between MTDMRG (blue) and DMRG-X (orange), for a~$3\times 3$ qubit array.}
\label{fig:mtdmrg_dmrgx_variance} 
\end{figure}

In~\cref{fig:mtdmrg_dmrgx_variance} we benchmark the wavefunction accuracy between DMRG-X and MTDMRG for the computation of the state corresponding to the highest frequency single-coupler excitation in a~$3\times 3$ qubit chip.

\section{\MakeUppercase{Circuit QED details}}
\label{app:circuiqed_details}

\subsection{Transmon qubit Hamiltonian}
\label{app:transmon_H}
\begin{figure}[t!]
\centering
\includegraphics[width=0.8\columnwidth]{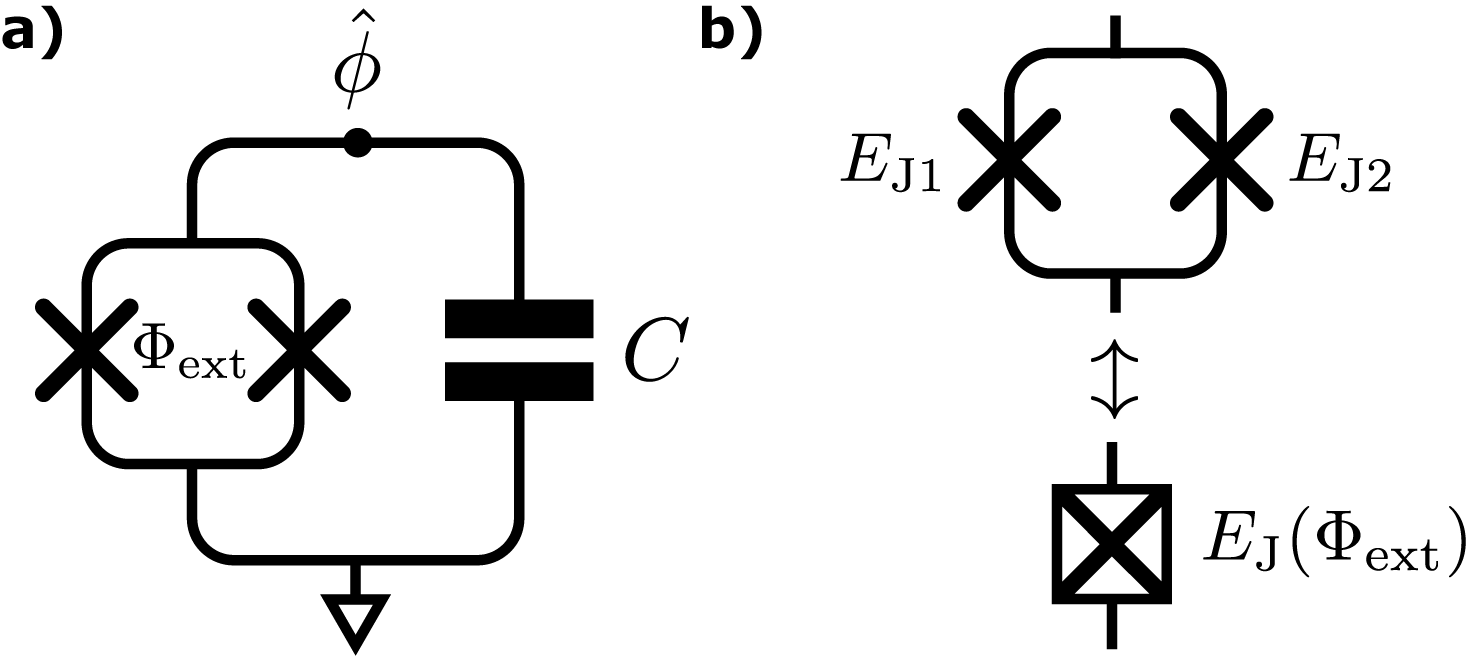}
\caption{\textbf{\textsf{a)}} Circuit diagram of a frequency-tunable transmon circuit, including a dc-SQUID with threaded magnetic flux~$\Phi_\mathrm{ext}$ (left) and the shunt capacitance~$C$ (right). \textbf{\textsf{b)}} The dc-SQUID is formed by two Josephson junctions arranged in parallel, with energies~$E_{\mathrm{J}1}$ and~$E_{\mathrm{J}2}$, leading to an equivalent flux-tunable Josephson element with Josephson energy~$E_\mathrm{J}(\Phi_\mathrm{ext})$. The threaded magnetic flux allows for tuning of~$E_\mathrm{J}(\Phi_\mathrm{ext})$, which determines the frequency of this superconducting qubit.}
\label{fig:transmon_diag} 
\end{figure}

In this section, we discuss a simplified version of the transmon Hamiltonian that we use in the main text. The transmon circuit is shown in~\cref{fig:transmon_diag}. Following~\cite{vool2017introduction}, we write down the Lagrangian, derive the Hamiltonian via a Legendre transformation, promote charge and flux variables to operators, and arrive at the description
\begin{equation}
    \hat H = \underbrace{4E_\mathrm{C} (\hat n -n_\mathrm{g})^2}_\text{kinetic energy}\underbrace{-E_\mathrm{J}(\Phi_\mathrm{ext})\cos\hat\phi}_\text{potential energy},
    \label{eq:transmon_hamiltonian_full}
\end{equation}
where~$E_\mathrm{C}=e^2/2C$ is the capacitive energy associated with the transmon capacitance~$C$, $n_\mathrm{g}$ is an offset-charge parameter, $E_\mathrm{J}(\Phi_\mathrm{ext})$ is the flux-tunable Josephson energy, and~$\hat n$ and~$\hat \phi$ are the reduced charge and phase operators of the transmon, satisfying~$[e^{\pm i\hat\phi},\hat n]=\mp e^{i\pm\hat\phi}$. The form of the flux-tunable Josephson energy follows from the potential energy of the circuit Hamiltonian in~\cref{fig:transmon_diag}a (which includes two Josephson junctions) after a gauge transformation---see Ref.~\cite{koch_2007} for the derivation. This transformation reduces the dc-SQUID to an effective flux-tunable Josephson element, shown in~\cref{fig:transmon_diag}b. 

\Cref{eq:transmon_hamiltonian_full} is supplemented with periodic boundary conditions for the wavefunction---i.e., the phase is compact~$\phi\in[-\pi,\pi)$~\cite{devoret2021does}. However, in the transmon regime~$E_\mathrm{J}(\Phi_\mathrm{ext})/E_\mathrm{C}\gg 1$, the potential energy dominates over the charging energy~\cite{gyenis2021moving}, and low-energy eigenstates~$|\psi_i\rangle$ of~\cref{eq:transmon_hamiltonian_full} display weak root-mean-square fluctuations of the phase~$\sqrt{\langle \psi_i|\hat\phi^2|\psi_i\rangle}\ll \pi$. In this limit, relying on a nonperiodic representation of the phase is a good approximation for low-lying eigenstates. Note that, as a result of this approximation, the sensitivity to the charge offset~$n_\mathrm{g}$ is lost~\cite{koch2009charging}, and we therefore set~$n_\mathrm{g}=0$ henceforth.

To arrive at a simplified model for~\cref{eq:transmon_hamiltonian_full}, we follow the review~\cite{blais2021circuit} as an excellent starting point. A more accurate mapping between the exact transmon Hamiltonian and the parameters of its Kerr model can be found in several places in the literature, see e.g., Refs.~\cite{didier2018analytical,petrescu2023accurate}.

The idea is to split~\cref{eq:transmon_hamiltonian_full} into linear and a nonlinear parts, i.e., $\hat H=\hat H_\mathrm{lin} + \hat H_\mathrm{nonlin}$, where
\begin{equation}
    \begin{split}
        \hat H_\mathrm{lin} &= 4E_\mathrm{C}\hat n^2 +\frac{E_\mathrm{J}(\Phi_\mathrm{ext})}{2}\hat\phi^2,\\
        \hat H_\mathrm{nonlin} &= -E_\mathrm{J}(\Phi_\mathrm{ext})\left(\cos\hat\phi + \frac{\hat\phi}{2}\right).
    \end{split}
    \label{eq:ham splitting}
\end{equation}
$H_\mathrm{lin}$ is diagonalized introducing the phase and charge quadratures 
\begin{equation}
    \begin{split}
    \hat\phi &=\sqrt{\xi}(\hat a +\hat a^\dagger),\\
    \hat n &=-\frac{i}{2\sqrt{\xi}}(\hat a -\hat a^\dagger), 
    \end{split}
    \label{eq:quadratures}
\end{equation}
defined in terms of the impedance-like parameter~$\xi=\sqrt{2E_\mathrm{C}/E_\mathrm{J}(\Phi_\mathrm{ext})}$, alongside the ladder-operator set~$(\hat a,\hat a^\dagger)$. Reintroducing~\cref{eq:quadratures} in~\cref{eq:ham splitting}, expanding the nonlinearity to 4th order, normal-ordering, and performing a rotating-wave approximation (where terms with unequal number of creation and annihilation operators are dropped), we arrive at the simplified Kerr model for the transmon device
\begin{equation}
    \hat H/\hbar \approx \omega_\mathrm{q}\hat a^\dagger \hat a -\frac{\eta_\mathrm{q}}{2}\hat a^{\dagger 2} \hat a^2,
\end{equation}
as stated in~\cref{sec:circuit QED}, with the frequency and anharmonicity parameters
\begin{equation}
    \begin{split}
    \hbar\omega_\mathrm{q}&\approx\sqrt{8 {E_{\mathrm{C}}}E_{\mathrm{J}}(\Phi_{\mathrm{ext}})}-E_{\mathrm{C}},\\
    \hbar\eta_\mathrm{q}&\approx E_{\mathrm{C}}. 
    \end{split}
\end{equation}

\begin{figure}[t!]
\centering
\includegraphics[width=0.7\columnwidth]{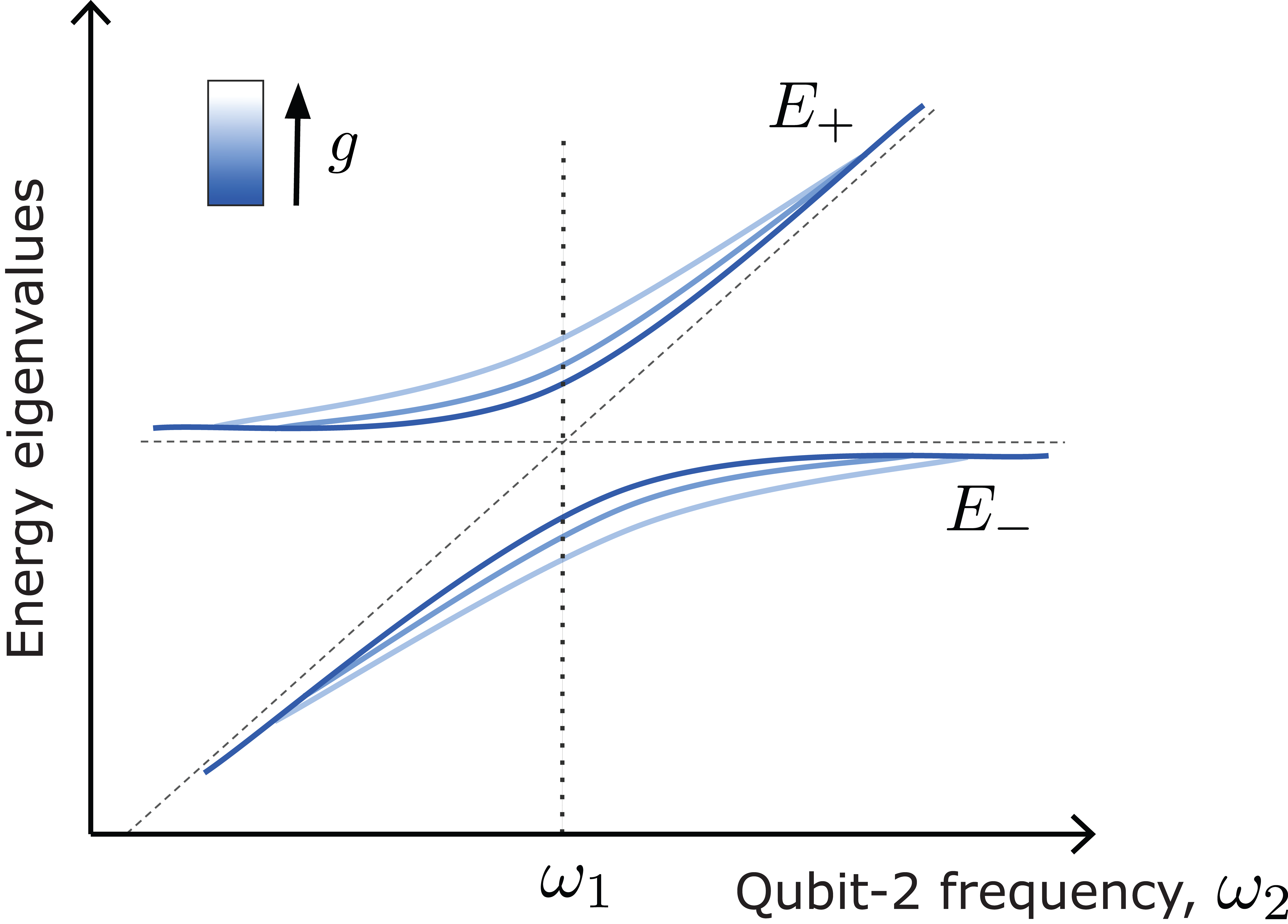}
\caption{Sketch of the avoided-level crossing between two nearly resonant qubits 1 and 2. The frequency of qubit 1, $\omega_1$, is kept fixed while the frequency of the second qubit, $\omega_2$, is swept around resonance. Energy eigenvalues are shown for different coupling strengths, depicting the minimum anticrossing (effective two-qubit coupling) becoming larger for larger coupling strengths.}
\label{fig:anticrossing_diag} 
\end{figure}

\subsection{Capacitive couplings}
\label{app:cap_couplings}

Here we describe the form of the capacitive-coupling Hamiltonian between two transmon modes. Our discussion avoids some of the technical details involved in the derivation of the full-circuit Hamiltonian. However, the interested reader can consult the reviews~\cite{blais2021circuit,krantz2019quantum} for more details.

In a processor, the individual transmon circuits, including qubits and couplers, are coupled via capacitances. Deriving the full-circuit Hamiltonian using a procedure similar to the one described in~\cref{app:transmon_H}, leads to the general form of the two-mode coupling
\begin{equation}
    \hat H_{ij} = 8 k_{ij}\sqrt{E_{\mathrm{C}i}E_{\mathrm{C}j}}\hat n_i \hat n_j,
    \label{eq:coupling_H_n}
\end{equation}
where~$\hat n_i$ and~$\hat n_j$ are the charge operators associated with transmons~$i$ and~$j$, respectively, $E_{\mathrm{C}i}$ and~$E_{\mathrm{C}j}$ are their capacitive energies [introduced in~\cref{eq:transmon_hamiltonian_full}], and 
\begin{equation}
    k_{ij} = \frac{C_{ij}}{\sqrt{C_i C_j}} + \mathcal{O}(C_{ij}^2),
    \label{eq:kij_def}
\end{equation}
is a coupling efficiency. The approximate sign in~\cref{eq:kij_def} holds in the limit of practical interest where the coupling capacitance~$C_{ij}$ is small in comparison to the mode capacitances, $C_i$ and~$C_j$.

Expressing the single-mode charge operators in terms of their bosonic expansion~\cref{eq:quadratures}, \cref{eq:coupling_H_n} takes the form
\begin{equation}
    \hat H_{ij} = \frac{8 k_{ij}\sqrt{E_{\mathrm{C}i}E_{\mathrm{C}j}}}{4\sqrt{\xi_i\xi_j}}(-i\hat a_i + i\hat a_i^\dagger)(-i\hat a_j + i\hat a_j^\dagger).
\end{equation}
Finally, implementing the gauge transformation~$\hat a_i\to i \hat a_i$ and~$\hat a_j\to i \hat a_j$, the coupling Hamiltonian reduces to
\begin{equation}
    \hat H_{ij}/\hbar = g_{ij}(\hat a_i + \hat a_i^\dagger)(\hat a_j + \hat a_j^\dagger),
\end{equation}
which agrees with~\cref{eq:coupling_hamiltonian} in the main text, and where the effective coupling strength between the two modes is
\begin{equation}
    \hbar g_{ij} = \frac{8 k_{ij}\sqrt{E_{\mathrm{C}i}E_{\mathrm{C}j}}}{4\sqrt{\xi_i\xi_j}}.
\end{equation}

\begin{figure}[t!]
\centering
\includegraphics[width=1\columnwidth]{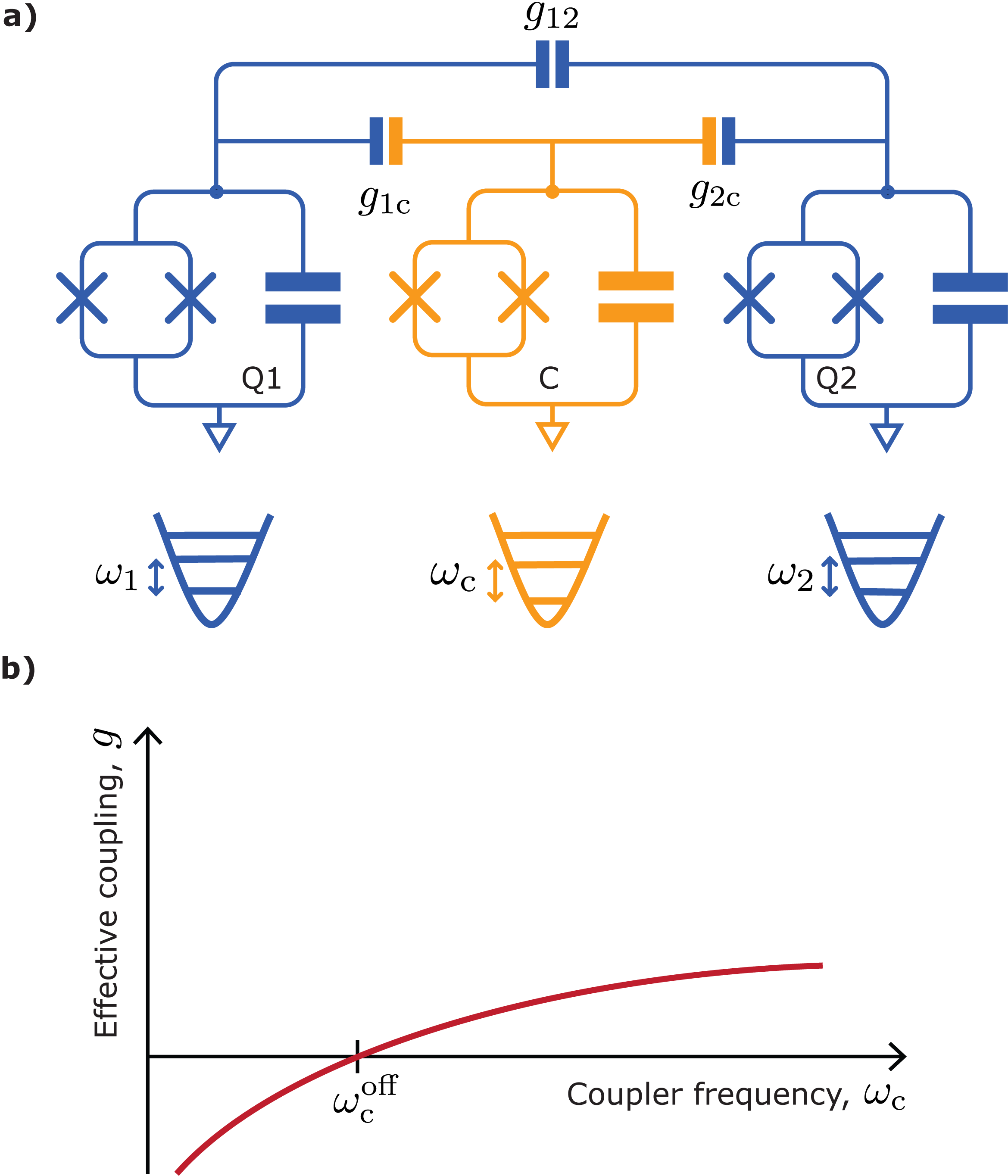}
\caption{\textbf{\textsf{a)}} Circuit diagram for a tunable coupler (C) between two qubits (Q1 and Q2). Each unit is a flux-tunable transmon. The qubits and coupler transitions frequencies are shown within a sketch of the low-energy portion of the Josephson potential for each mode. The transmons are coupled capacitively. \textbf{\textsf{b)}} Sketch of the effective coupling between the two qubits (blue) as a function of the frequency of the intermediary transmon coupler (orange).}
\label{fig:tunable_coupler} 
\end{figure}

\subsection{Two-qubit avoided level crossing}
\label{app:xy_splitting}

The bosonic Hamiltonian for two interacting transmons as presented in~\cref{sec:circuit QED} is:
\begin{equation}
    \hat H_i/\hbar = \sum_{i=1,2} \omega_i\hat a_i^\dagger \hat a_i -\frac{\eta_i}{2}\hat a_i^{\dagger 2} \hat a_i^2 + g(\hat a_1 + \hat a_1^\dagger)(\hat a_2 + \hat a_2^\dagger).
    \label{eq:two_mode_ham}
\end{equation}
Near the resonance condition~$\omega_1\approx\omega_2$, the Hamiltonian restricted to the single-excitation subspace is
\begin{equation}
    \hat{H}_{\{01,10\}}/\hbar \approx \begin{pmatrix}
    \omega_1 & g \\
    g & \omega_2
  \end{pmatrix}.
\end{equation}
The two energy eigenvalues of this Hamiltonian are
\begin{equation}
    E_{\pm}/\hbar = \frac{\omega_1 + \omega_2}{2} \pm \sqrt{\frac{\delta^2}{4} + g^2},
\end{equation}
with~$\delta = \omega_1-\omega_2$, representing an avoided-level crossing as a function of detuning~$\delta$, see~\cref{fig:anticrossing_diag}. For~$\delta = 0$, the energy levels experience their minimal splitting, $\Delta = (E_{+}-E_{-})/\hbar = 2 g$, where~$g$ is a measure of the two-qubit coupling in~\cref{eq:two_mode_ham}. In a more general setting with more transmons, the minimal energy splitting corresponds to the effective (dressed) coupling between two modes.

\begin{figure}[t!]
\centering
\includegraphics[width=0.7\columnwidth]{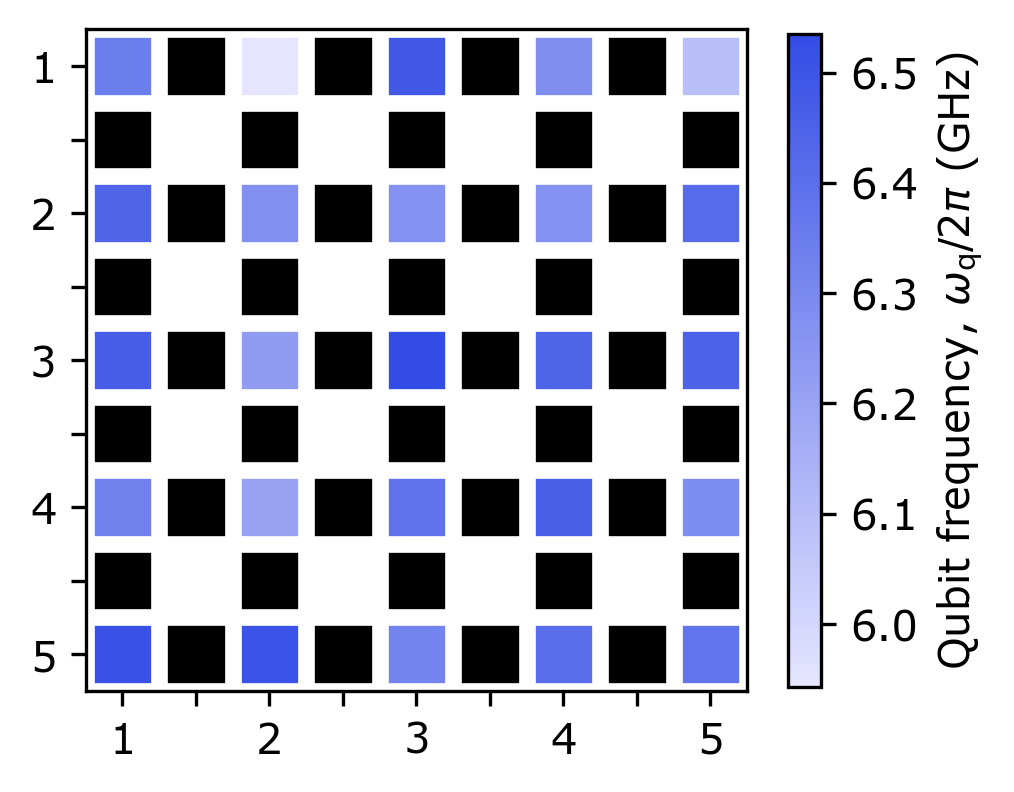}
\caption{Bare qubit frequencies selected for the study in~\cref{sec:qubit_coupler_loc}. The coupler frequencies have been optimized to suppress the single-excitation coupling~$g$ between neighboring qubits.}
\label{fig:loc_exp_qfrequencies} 
\end{figure}

\subsection{Tunable-coupler scheme}
\label{app:tunable_coupler}

The qubit architecture that we consider incorporates a tunable coupling scheme~\cite{neill2017path, Yan_2018}. That is, the transmons that encode the qubit degrees of freedom are connected by couplers that are, themselves, transmons. 

The circuit diagram for the two-qubit+coupler system is depicted in~\cref{fig:tunable_coupler}a. The effective coupling, $g$, between the two qubits is a function of the coupler frequency (\cref{fig:tunable_coupler}b). Provided the direct qubit-qubit and qubit-coupler couplings are chosen appropriately, for every pair of transmon qubits coupled by a coupler transmon, there exists a coupler frequency~$\omega_c^{\text{off}}$ such that the effective two-qubit interaction is minimized, ideally approaching zero. This defines the noninteracting (idle) operation regime of the device. To enact gates, the coupler is biased away from the `off' point, turning `on' an exchange interaction between the qubits. 

\section{\MakeUppercase{Numerical simulation details}} \label{app:numerical_details}

The bond dimension used in the qubit and coupler localization numerical experiment (\cref{sec:qubit_coupler_loc}) is~$\chi = 80$ for the DMRG-X calculation of each single transmon excitation state. This yielded wavefunction variances in the range~$10^{-7} - 10^{-9}$ (GHz$^2$). The local Hilbert space dimension for qubits is~$d_\mathrm{q}=4$ and for couplers it is~$d_\mathrm{c}=3$. The bare frequencies used to construct the Hamiltonian and corresponding MPO for the qubits in this study are depicted in~\cref{fig:loc_exp_qfrequencies}.\\

In the MTDMRG-X simulation in~\cref{sec:results_mtdmrgx}, the bond dimension used for the MTMPS was~$\chi = 20\times|S|$ (where~$|S|$ is the number of reference bare states). This yields wavefunction variances in the range~$10^{-10}-10^{-8}$ (GHz). The local Hilbert space dimension for qubits (couplers) was~$d_\mathrm{q}=4$ ($d_\mathrm{c}=3$).

\end{document}